\documentclass[fleqn,usenatbib]{mnras}

\usepackage{newtxtext,newtxmath}

\usepackage[T1]{fontenc}

\DeclareRobustCommand{\VAN}[3]{#2}
\let\VANthebibliography\thebibliography
\def\thebibliography{\DeclareRobustCommand{\VAN}[3]{##3}\VANthebibliography}

\usepackage{graphicx}	
\usepackage{amsmath}	

\usepackage{psfrag,color}
\usepackage{color,soul} 
\usepackage{booktabs}
\usepackage{graphicx}	
\usepackage{amsmath}	

\usepackage[utf8]{inputenc}

\newcommand{\Ne}{{$n_{\rm e}$}}
\newcommand{\Te}{{$T_{\rm e}$}}
\newcommand{\TO}{{$T_{\rm e}$[\ion{O}{iii}]}}
\newcommand{\To}{{$T_{\rm e}$[\ion{O}{ii}]}}
\newcommand{\TS}{{$T_{\rm e}$[\ion{S}{iii}]}}
\newcommand{\TN}{{$T_{\rm e}$[\ion{N}{ii}]}}

\newcommand\W{{$\lambda$}}
\def\i{\,{\sc i}}

\def\v{\,{\sc v}}

\newcommand{\hii}{H\thinspace{\sc ii}}

\usepackage{orcidlink}

\title[Nitrogen across cosmic time]{CLASSY XII: Nitrogen Enrichment Shaped by Gas Density and Feedback}

\author[K. Z. Arellano-C\'ordova]{
K. Z. Arellano-C\'ordova \orcidlink{0000-0002-2644-3518} $^{1}$\thanks{E-mail: ziboney@gmail.com and k.arellano@ed.ac.uk (KZAC)},
D. A. Berg \orcidlink{0000-0002-4153-053X}$^{2}$,
M. Mingozzi \orcidlink{0000-0003-2589-762X}$^{3}$,
B. L. James \orcidlink{0000-0003-4372-2006}$^{3}$,
F. Vincenzo \orcidlink{0000-0002-0743-9994}$^{4}$\and\
N. S. J. Rogers \orcidlink{0000-0002-0361-8223}$^{5}$,
E. D. Skillman \orcidlink{0000-0003-0605-8732}$^{6}$,
R. O. Amor\'{i}n    \orcidlink{0000-0001-5758-1000}$^{7}$,
F. Cullen       \orcidlink{0000-0002-3736-476X}$^{1}$,
S. R. Flury     \orcidlink{0000-0002-0159-2613}$^{1}$,\and\
V. Abril-Melgarejo \orcidlink{0000-0002-2764-6069}$^{3, 8}$,
J. Chisholm \orcidlink{0000-0002-0302-2577}$^{2}$,
T. Heckman \orcidlink{0000-0000-0000-0000}$^{9}$,
M. J. Hayes \orcidlink{0000-0001-8587-218X}$^{10}$,
S. Hernandez \orcidlink{0000-0003-4857-8699}$^{3}$, \and\
N. Kumari \orcidlink{0000-0002-5320-2568}$^{3}$,
C. Kobayashi \orcidlink{0000-0002-4343-0487}$^{11}$,
C. Leitherer \orcidlink{0000-0000-0000-0000}$^{3}$,
C. L. Martin \orcidlink{0000-0001-9189-7818}$^{12}$,
Z. Martinez \orcidlink{0009-0000-2997-7630}$^{2}$ 
T. Nanayakkara \orcidlink{0000-0003-2804-0648}$^{13}$ \and\
K. S. Parker \orcidlink{0000-0002-8809-4608}$^{2}$,
P. Senchyna \orcidlink{0000-0002-9132-6561}$^{14}$,
C. Scarlata \orcidlink{0000-0002-9136-8876}$^{6}$, 
M. G. Stephenson \orcidlink{0000-0000-0000-0000}$^{2}$,
A. Wofford \orcidlink{0000-0001-8289-3428}$^{15}$,
X. Xu \orcidlink{0000-0002-9217-7051}$^{5}$
P. Zhu \orcidlink{0000-0002-1333-147X}$^{16, 17, 18}$\\
$^{1}$ Institute for Astronomy, University of Edinburgh, Royal Observatory, Edinburgh EH9 3HJ, UK \\
$^{2}$ Department of Astronomy, The University of Texas at Austin, 2515 Speedway, Stop C1400, Austin, TX 78712, USA\\
$^{3}$AURA for ESA, Space Telescope Science Institute, 3700 San Martin Drive, Baltimore, MD 21218, USA\\
$^{4}$ Dipartimento di Fisica e Astronomia ‘Ettore Majorana’, Università degli Studi di Catania, Via S. Sofia 64, 95123 Catania, Italy\\
$^{5}$ Center for Interdisciplinary Exploration and Research in Astrophysics (CIERA), Northwestern University, 1800 Sherman Ave., Evanston, IL, 60201, USA\\
$^{6}$ Minnesota Institute for Astrophysics, University of Minnesota, 116 Church Street SE, Minneapolis, MN 55455, USA\\
$^{7}$ Instituto de Astrof\'{i}sica de Andaluc\'{i}a (CSIC), Apartado 3004, 18080 Granada, Spain\\
$^{8}$ LUX, Observatoire de Paris, Université PSL, CNRS, 5 place Jules Janssen, F-92190 Meudon, France\\
$^{9}$ Center for Astrophysical Sciences, Department of Physics \& Astronomy, Johns Hopkins University, Baltimore, MD 21218, USA\\
$^{10}$ Stockholm University, Department of Astronomy and Oskar Klein Centre for Cosmoparticle Physics, AlbaNova University Centre, SE-10691, Stockholm, Sweden\\
$^{11}$ Centre for Astrophysics Research, Department of Physics, Astronomy and Mathematics, University of Hertfordshire, Hatfield, AL10 9AB, UK\\
$^{12}$ Department of Physics, University of California, Santa Barbara, Santa Barbara, CA 93106, USA\\
$^{13}$ Swinburne University of Technology, Melbourne, Victoria, AU\\
$^{14}$ Carnegie Observatories, 813 Santa Barbara Street, Pasadena, CA 91101, USA\\
$^{15}$ Instituto de Astronom\'{i}a, Universidad Nacional Aut\'{o}noma de M\'{e}xico, Unidad Acad\'{e}mica en Ensenada, Km 103 Carr. Tijuana-Ensenada, Ensenada 22860, M\'{e}xico\\
$^{16}$ Research School of Astronomy and Astrophysics, Australian National University, Australia\\
$^{17}$ ARC Centre of Excellence for All Sky Astrophysics in 3 Dimensions (ASTRO 3D), Australia \\
$^{18}$ Center for Astrophysics | Harvard \& Smithsonian, 60 Garden Street, Cambridge, MA 02138, USA\\
}
\date{Accepted XXX. Received YYY; in original form ZZZ}

\pubyear{2015}

\begin{document}
\label{firstpage}
\pagerange{\pageref{firstpage}--\pageref{lastpage}}

\maketitle

\begin{abstract}
We investigate the chemical evolution of N/O using a sample of 45 local star-forming galaxies (SFGs) from the CLASSY survey. This sample spans a wide range of galaxy properties, with robust determinations of nitrogen and oxygen abundances via the direct-$T_{\rm e}$ method. We explore how N/O relates to density structure, stellar mass, star formation rate (SFR), stellar age, compactness, and gas kinematics. In addition, we compare our results with those of galaxies at $z =2-10$ where N/O ratios were derived from optical or UV nitrogen lines, aiming to identify chemical enrichment pathways across cosmic time. Our analysis shows that the N/O-O/H relation in CLASSY galaxies aligns with the trends seen in local galaxies and extragalactic \ion{H}{ii} regions, and that galaxies at $z = 2-6$ exhibit similar N/O values, indicating no significant redshift evolution in N/O for a fixed metallicity. We identify a significant correlation between electron density $n_{\rm e}$([\ion{S}{ii}]) and N/O, suggesting that density structure contributes to the scatter in the N/O-O/H relation. The CLASSY galaxies with high SFRs or compact star formation show elevated N/O, though no strong correlation with stellar mass is found. We also find that high-velocity outflows  ($v_{\rm out}\ga 350$ km/s) and low mass-loading factors are linked to elevated N/O, indicating that feedback plays a significant role. These results highlight the importance of density, star formation, and feedback from young stellar populations in shaping N/O enrichment and provide key insights for interpreting high-$z$ galaxies observed with JWST.
\end{abstract}




\section{Introduction}

Understanding the star-formation history of galaxies requires a detailed study of stellar mass assembly across redshifts. A fundamental probe of galaxy evolution is the chemical enrichment traced by the nucleosynthesis products of generations of stars, which provide insight into the mechanisms associated with the baryon cycle. 

The abundances of carbon (C), nitrogen (N), and oxygen (O) relative to hydrogen (H) are among the most important tracers of the chemical enrichment in star-forming regions across cosmic time. 
O is produced over very short timescales ($\mathrm{\sim 10\,Myr}$) in massive stars and is subsequently ejected into the interstellar medium (ISM) by core-collapse supernovae (CCSNe). The oxygen abundance is commonly used as a proxy for the gas-phase metallicity. N is produced during the CNO cycle and is ejected into the ISM by both massive and intermediate-mass stars $\mathrm{4- 7}$M$_{\odot}$ at a timescale of ($\mathrm{\sim 250 Myr}$), and continues to be ejected over $\mathrm{1 Gyr}$. C is also produce by massive and low-mass stars ($\mathrm{1-4}$M$_{\odot}$), and serves to catalyzer for the production of N. \citep[e.g.,][]{matteucci86, henry00, ventura13, vincenzo16, Schaefer20, Johnson23,Rizzuti25, kobayashi11}. 
These elements are released on different timescales through supernova explosions and violent stellar winds, expelling C, N, and O into the ISM, where they become subject to the complex processes of the baryon cycle. For example, inflows of pristine gas dilute the metallicity and fuel new star formation, while outflows and gas recycling remove or redistribute metals~\citep[e.g.,][]{tumlinson17}.

In this context, the different production timescales of nitrogen and oxygen make it possible to build the N/O-O/H relation, which is an essential diagnostic of star formation and chemical enrichment histories of galaxies.
The shape of the N/O-O/H relation mainly reflects the primary and secondary origins of N, determined by initial C production through triple-$\alpha$ burning and by pre-existing C from the CNO cycle, respectively. At metallicities ($\mathrm{12+log(O/H)\la 8}$), the chemical enrichment of N/O remains constant with O/H ("plateau") indicating that N is evolving in lockstep with O, which is associated with the only contribution of massive stars. As metallicity increases, the AGB stars eject N into the ISM, increasing the N/O ratio \citep[e.g.,][]{henry00, nava06, perez-montero09, vincenzo16, kobayashi20, Johnson23, kobayashi11}.

In addition to the nucleosynthesis pathways of N and O, there are other physical mechanisms that shape the N/O-O/H ratio. For example, such a scatter might be related to differences in star formation histories, and low star-formation efficiency that reduces the production of oxygen while the production of N still continues from more evolved stars, and the chemical enrichment of N via the strong winds of  Wolf-Rayet stars (WR). Another important scenario that might contribute to the scatter in the N/O-O/H relation is mechanical feedback from massive stars, which can transport metals from stellar production sites out into the ISM \citep[e.g.,][]{Chisholm18}. At low metallicity, supernovae (SNe) can be suppressed below the typically assumed threshold \citep[e.g.,][]{jecmen23}, which in turn can stunt the production of O via CCSNe. The lower O yields from such low metallicity stellar populations would result in a net increase in N/O. Additionally, gas inflows can also play an important role in the N/O ratios, since inflows of pristine gas dilute the metallicity without altering nitrogen, and hence increase N/O \citep[e.g.,][]{edmunds01, koppen&hensler05, amorin10, perezdiaz24, kobayashi24}.

Local star-forming galaxies (SFGs) and extragalactic \ion{H}{ii} regions are a key laboratory for demonstrating which physical mechanisms and conditions drive the chemical abundance patterns of N and O across cosmic time. In this context, local SFGs offer an opportunity to study the chemical enrichment of N and O in much detail to disentangle the different gas conditions and galaxy properties that might contribute to the observed scatter N/O-O/H relation and that cannot be explained by the standard nucleosynthesis pathways of N/O. 

In principle, the total abundance of an element is the contribution of each ionic species relative to hydrogen (e.g., N$^{+}$/H$^{+}$, N$^{2+}$/H$^{+}$ and N$^{3+}$/H$^{+}$), and its contribution will depend on the ionization state of the gas as well as the physical conditions of the ionized gas. The low ionization lines of [\ion{N}{ii}]~\W\W 6548,84 are the only collisional excited lines of N available in the optical, with a more that 40\% in their ionic abundance fraction of N$^{+}$/N contributing to the total N/H in low ionization objects \citep[e.g.,][and Martinez et al. 2025.]{amayo21, berg21a}. In optical, the N/O abundance ratio is derived mostly using the [\ion{N}{ii}]~\W\W 6548,84 and [\ion{O}{ii}]~\W\W3726,29 lines (or if available  [\ion{O}{ii}]~\W\W7320,7330 but these lines are less commonly used) in \ion{H}{ii} regions and SFGS \citep[e.g.,][]{perez-montero09, perezmontero13,croxall16, berg12, arellanocordova20, stephenson23, Henry06}. If $T_{\rm e}$ is not available, N/O can also be estimated using strong-line methods, such as the [\ion{N}{ii}]/[\ion{O}{ii}] and [\ion{N}{ii}]/[\ion{S}{ii}] ratios. \citep{perezmontero17, florido22, hayden-pawson22}.

With the several detections of UV N lines, \ion{N}{iii}] \W\W1750 and \ion{N}{iv}] \W1485 by the {\it JWST}, the higher ionization states of nitrogen can also be traced in high-$z$ SFGs, allowing the opportunity to analyze the gas properties and chemical enrichment of N. Overall, measurements of N/O at high-$z$ suggest that a significant portion of galaxy populations at these redshifts is characterized by extremely large N/O at low metallicity \citep[e.g.,][]{Bunker23, Cameron23, isobe23a, marqueschaves24, schaerer24, curti24b, castellano24, hayes25}.
This elevated N/O might be explained by different scenarios such as the presence of massive stars or very massive stars, variations of the initial mass function (IMF), chemical enrichment of N/O via WR stars, gas heating by shocks, and the presence of active galactic nuclei \citep[e.g.,][]{senchyna24, vink23, charbonnel23, watanabe24, Maiolino24, flury24}.
Recent studies reveal that globular clusters share the same N/O enrichment observed in high-$z$ galaxies, where dense environments enabled the formation of very massive stars, pointing to a possible common origin in the early universe \citep[e.g.,][]{senchyna24, Ji25}. Alternatively, studies using chemical evolution modeling concluded that high UV N/O can be explained via different episodes of star-formation following a quiescent phase \citep{kobayashi24}. 
However, it is still unclear what is the impact in the abundances interpretation of using UV high-ionization lines and optical low-ionization lines tracing the N/O abundances.

In this work, we investigate the enrichment pathways of nitrogen and oxygen in the N/O-O/H relation by using 45 local SFGs from the COS Legacy Archive Spectroscopic SurveY (CLASSY) \citep[e.g.,][]{berg22, james22}. The CLASSY galaxies have well-constrained direct abundances, total stellar masses (M$_\star$), star formation rates (SFRs), and gas physical and chemical properties derived from UV and optical data \citep[e.g.,][]{berg22, james22, mingozzi22, xu22}. Our sample of local SFGs mimics the physical properties of high-$z$ galaxies, particularly due to their high specific star-formation rates. Therefore, the wide range of physical properties in CLASSY enables us to study how these galaxy properties impact the observed N/O and N/H ratios. In particular, since most CLASSY galaxies are undergoing strong star formation, we can trace variations in N/O across different metallicities. With these results at $z\sim0$, we can shed light on the chemical enrichment that shapes the N/O-O/H relation at all metallicities by addressing what properties are implied on the observed scatter at $z\sim0$, and if such scenarios might be extended to high-$z$.

The structure of the paper is as follows: 
In Sec.~\ref{sec: sample}, we present the CLASSY galaxies and the archival properties analyzed here,  and an additional high-redshift sample gathered from the literature for comparison with our results. Sec.~\ref{sec:physical_conditions} describes the methodology used to derive the physical conditions and chemical abundances of N and O of this study. 
In Sec.~\ref{results}, we present the N/O-O/H  and N/H-O/H relations for CLASSY . We analyze the role of electron density, SFR and kinematics in the observed scatter in the N/O-O/H relation.  In these sections, we also compare our results with those of the high-$z$ galaxies, and we present the results of the SFR and stellar mass surface densities for the CLASSY sample. Moreover, we introduce new chemical evolution models to explore the scatter of the N/O-O/H relation at low-metallicities. The discussion of our results and the comparison of N/O across redshifts in Sec~\ref{sec:discussion}. Finally, Sec~\ref{sec:conclusion} summarizes our conclusions.

In this paper, we use the following solar abundance ratios taken from \citet{asplund21}: 12+log(O/H) = 8.69$\pm$0.04 and 12+log(N/H) = 7.83$\pm$0.07.  We adopt a flat $\Lambda$CDM cosmology with $\Omega_m = 0.3$, $\Omega_\Lambda = 0.7$, and $H_0 = 70\,\mathrm{km\,s^{-1}\,Mpc^{-1}}$.

\section{Sample}\label{sec: sample}
\subsection{CLASSY ($z\sim0)$}
Our sample comprises 45 optical spectra of SFGs ($0.002<$ $z$ $<0.182$) mainly gathered from the Sloan Digital Sky Survey \citep[SDSS][]{abazajian09}, with six objects from other facilities such as the Large Binocular Telescope (LBT)/MODS, and the Very Large Telescope/MUSE/VIMOS. These observations are part of the CLASSY treasury survey \citet[][]{berg22}. A general overview of CLASSY can be found in \citet{berg22}, and \citet{james22}, while the galaxy properties are reported in \citet[][]{berg22}, \citet{mingozzi22} and \citet{xu22}. Overall, CLASSY covers a broad range of physical properties in terms of stellar mass, SFR , metallicity, and ionization parameter. However, note that CLASSY is biased towards UV-bright ($m_{\rm FUV} < 21$ AB arcsec$^{-2}$) and high SFR with the goal of mimicking high$-z$ galaxies. 
In \citet[][hereafter AC24a]{arellanocordova24}, we presented a study of the chemical abundance patterns of Ne, S, Cl, and Ar with O. Here, we continue that study by analyzing the N abundances. We follow the methodology described in AC24a, which we summarize in Sec.~\ref{sec:physical_conditions}.

\subsection{Compilation of galaxy properties}\label{sec_sub:gal_properties_comp}
Additionally, we used previous results from CLASSY to investigate the relationship between N/O and various gas and galaxy properties across different scaling relations. In particular, we considered the color excess ($E(B-V)$), stellar mass, SFR, and optical sizes from \citet{berg22}. The $E(B-V)$ values were derived from the Balmer decrement ratios, using the extinction law of \citet{cardelli89}. These $E(B-V)$ values were applied to correct the observed fluxes of different ions for dust attenuation, as reported by \citet{mingozzi22}. We also collected the equivalent width (EW) of H$\beta$ from \citet{mingozzi22}.

For this study, we selected the set of galaxy properties that characterize the entire galaxy \citep[see][for details]{berg22}. The stellar masses and SFRs were derived from spectral energy distribution (SED) fitting, assuming a constant star formation history and a \citet{chabrier03} IMF. The optical sizes of the galaxies (half-light radius, $r_{50}$) were measured using imaging in various bands, such as SDSS and Pan-STARRS, as reported in \citet[][see their Table 6]{berg22}. Finally, we also analyzed the impact of stellar population age on the N/O results. We compiled stellar ages for the CLASSY sample from Parker et al. in preparation, inferred using the method of \citet{chisholm19}. This method consists of modeling the UV stellar continuum with a single-age stellar population from {\tt Starburst99} \citep{leitherer99}, assuming a Kroupa IMF \citep[see][and Parker et al. in preparation]{chisholm19}.

\subsection{Archival Sample at high-$z$}\label{sec:Archival_data}
In this analysis, we have incorporated the N/O and O/H data from the sample of \hii\ regions and local SFGs extracted from the literature and presented in \citet{arellanocordova24}.  This sample corresponds to the work of \citet{berg19a, izotov06, izotov17, izotov21d, james15, Rogers22}. We have also added the sample of Lyman Break Galaxies from \citet[][$z<0.2$]{Loaiza-Agudelo20}, whose chemical properties were recalculated here using the dereddened fluxes reported by the authors.

 We have compiled a sample of 22 galaxies at $z=2-10$ with direct abundance determination where N/O is derived using either optical or UV N emission lines. This high-redshift sample with N/O from the optical lines includes: two galaxies at $z\sim2$ from \citet{sanders23}, The Sunburst Arc at $z =2.37$ from \citet{welch25}, Q2343-D40 from \citet{rogers24} at $z=2.96$, two $z\sim5$ EXCELS galaxies from \citet{arellanocordova25} and 10 EXCELS galaxies from \citet{scholte25} and \citet{stanton2025}. For the sample with N/O derived using UV lines (e.g., [\ion{N}{iv}] and \ion{N}{iii}]),  we consider $z > 6$ galaxies from \citet[][RXCJ2248-ID, $z=6.11$]{topping24}, \citet[][GLASS$\_$15008, $z=6.23$]{isobe23a}, \citet[][A1703-zd6, $z=7.04$]{topping24b}, \citet[][CEERS$\_$1019, $z=8.63$]{marqueschaves24}, \citet[][GN-z9p4, $z=9.38$]{schaerer24}, and \citet[][GS-z9-0, $z=9.43$]{curti24b}. 
 We have distinguished the high redshift sources according to the N emission lines used for calculating N/O. In the following figures of this study, the SFGs that use optical [\ion{N}{ii}] $\lambda$6584 to derive N/O are represented by pentagons, while the sample that uses UV [\ion{N}{iv}] or \ion{N}{iii}] to derive N/O is represented by squares. It is worth mentioning that $z > 6$ objects may have a different nature compared to the CLASSY galaxies, as they are extremely compact clumps. However, we decided to include them in order to show the parameter space they occupy in the N/O-O/H relation and in other scaling relations presented in this study.

\section{Physical Conditions and Chemical abundances }\label{sec:physical_conditions}
\subsection{Electron density and temperature}
For our analysis of N and O, we assumed the same electron temperature (\Te) and density ($n_e$) structure for each galaxy presented in AC24a. These authors calculated electron densities using the [\ion{S}{ii}] \W6731/\W6717 density diagnostic and a three-zone temperature diagnostic, tracing the low (\To), intermediate (\TS), and high (\TO) ionization emitting regions of the gas. These temperatures were derived using the following diagnostics: [\ion{O}{ii}] (\W\W3726,3729)/(\W\W7319,7320 \W\W7330,7331)\footnote{Hereafter referred to as [\ion{O}{ii}] \W3727 and [\ion{O}{ii}] \W\W7320,7330 since these lines are blended due to the spectral resolution of the sample}, [\ion{N}{ii}]~(\W\W6548,6584)/\W5755, [\ion{S}{iii}]~\W\W9069,9532)/\W6312, and [\ion{O}{iii}]~(\W\W4959,5007)/\W4363, respectively.  
The calculations of the electron density and temperature were done using the nebular analysis package PyNeb \citep[][version 1.1.14]{luridiana15}  and the atomic data reported in Table~\ref{tab:atomic_data}.

 \begin{table*}\footnotesize
 \caption{Atomic data used in this work}
 \begin{center}
 \begin{tabular}{lcc}
 \hline
 \multicolumn{1}{l}{Ion} & \multicolumn{1}{c}{Transition Probabilities (A$_{ij}$)} & \multicolumn{1}{c}{Collision Strengths ($\Upsilon_{ij}$)} \\
 \hline

O$^{+}$   &  \citet{fft04}   & \citet{Kisielius:2009} \\
O$^{2+}$  &  \citet{fft04}  & \citet{aggarwal99} \\
 N$^{+}$   &  \citet{fft04}  & \citet{tayal11}  \\
S$^{+}$   &  \citet{Podobedova:2009} &  \citet{Tayal:2010} \\
S$^{2+}$  &  \citet{Podobedova:2009}  &  \citet{Grieve:2014} \\



 \hline
 \end{tabular}
 \end{center}
 \label{tab:atomic_data}
 \end{table*}

  %
\subsection{Ionic and total abundances of N and O}
To derive accurate ionic abundances, it is important to consider the appropriate \Te~ structure of each ionic species based on the ionization potential. \citet{mendez-delgado23b} showed that the use of \To\ can overestimate the ionic abundances, since \To\ strongly depends on the electron density.
To evaluate the impact of using \TN\ and \To\ in calculating the N$^{+}$/O$^{+}$ ratio, we have selected 13 SFGs from CLASSY with measurements of both \TN\ and \To. The electron density for these galaxies ranges from 100 cm$^{-3}$ to 480 cm$^{-3}$ (for comparison between \To\ and \TN\ see Fig. 8 in AC24a)

\begin{figure}
\begin{center}
    \includegraphics[width=0.45\textwidth, trim=30 0 30 0,  clip=yes]{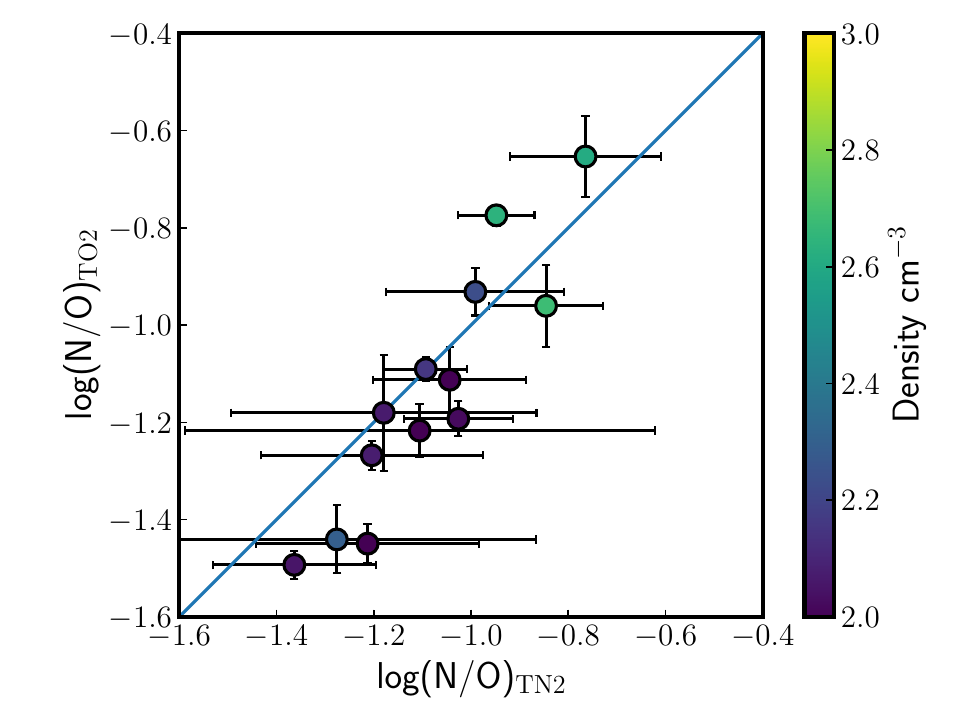}
        \caption{Comparison between log(N/O) derived using \To\ and log(N/O) derived using \TN, labeled as TO2 and TN2, respectively. The solid line represents the 1:1 relation, and the results are color-coded with log(\Ne[\ion{S}{ii}]). Note that the uncertainties in the measurements of log(N/O) are larger when using \TN\ compared to \To. It  shows that the two values are consistent within the uncertainties, with a slight offset toward lower values of log(N/O) at low \Ne when using \TN.} 
\label{fig:no_to2tn}
\end{center}
\end{figure}

In Fig. \ref{fig:no_to2tn}, we present the results of the N/O derived using \To\ and \TN. For these 13 CLASSY galaxies, both results are consistent within the uncertainties, with differences $<$ 0.2 dex. The figure also illustrates that galaxies with high densities tend to have higher values of N/O. We discuss this in more detail in Sec~\ref{sec:density}.
We note that the N/O ratios using \TN\ are more uncertain due to the faintness of the \Te-sensitive [\ion{N}{ii}] \W5755. 
A larger sample of ionized objects with both \To\ and \TN\ is needed to corroborate the results in Figure~\ref{fig:no_to2tn}. Since \To\ depends on density, and given the limited sample of galaxies with \TN\ measurements, we derive the ionic abundances of O$^{+}$ and N$^{+}$ using an estimated \Te\ from \TO, based on the temperature relations of \citet{garnett92} \citep[see also][]{arellanocordova20, perezmontero17}. For galaxies with only one measurement of \To\ or \TN\ and without \TO, we use those temperatures to estimate the low-ionization ionic abundances of N and O.

For the high-ionization ion O$^{++}$, we use \TO\ in 33 galaxies. For the remaining 10 galaxies, we use either \TS\ or \To\ to estimate the high-ionization \Te, employing the temperature relations from \citet{garnett92} and \citet{Rogers21}. The results of the temperature structure are consistent with those reported by AC24a (see their Table~5). Therefore, with the two different ionic species of O available, we infer the metallicity of the CLASSY galaxies by adding the contributions from O$^{+}$/H$^{+}$ and O$^{2+}$/H$^{+}$.

To calculate the total abundances of N/O and N/H, we use ionization correction factors (ICFs) to account for unobserved transitions of different ionization species. ICFs are particularly important because they are one of the major sources of uncertainty in chemical abundance determinations. Following AC24a, we analyzed different ICFs to select the most appropriate for this analysis. We derived N/O abundances using the ICF proposed by \citet{peimbert69}, N$^{+}$/O$^{+}$ $\approx$ N/O, based on the similar ionization potentials of O$^{+}$ and N$^{+}$. This ICF is widely used in \hii\ regions and SFGs \citep[e.g.,][]{croxall16, esteban20, arellano-cordova2020b, berg20, stephenson23}.

Since the ICF of \citet{peimbert69} is often chosen for N/O studies in the literature, we compare it with
the ICFs from \citet{izotov06} and \citet{amayo21}, which depend on the ionization parameter defined as O$^{2+}$/(O$^{+}$ + O$^{2+}$). We find consistent results between the ICFs of \citet{peimbert69} and \citet{izotov06}, with mean differences of 0.003 dex and a standard deviation of $\sigma = 0.04$ dex. However, the ICF of \citet{amayo21} shows a systematic shift to higher values of N/O with respect to the ICFs of \citet{izotov06} and \citet{peimbert69}, but has a significantly lower dispersion than the others. The differences in N/O between \citet{amayo21} and \citet{peimbert69} have means and standard deviations of 0.07 dex and 0.11 dex, respectively. We have adopted the ICF of 
\citet{peimbert69} to provide the N/O and N/H abundances for CLASSY. The results of the ionic and total abundances for the N/H and N/O ratios are reported in Table~\ref{tab:ionic_sfr_mass_density_classy}. 

\begin{table*}
\caption{ Electron densities ($n_{\rm e}$[\ion{S}{ii}] and $n_{\rm e}$[\ion{Ar}{iv}]), total oxygen abundance, and total nitrogen abundance for CLASSY galaxies are listed in columns 2-6. Stellar mass surface density and SFR surface density are reported in columns 7-8 (see Sec.\ref{sec:surface_Density_analysis}). Column 9 indicates references for CLASSY galaxies previously identified with WR features. SFGs showing \ion{He}{ii}~\W4686 emission with FWHM $\ga$ 500 km/s are considered WR candidates. }
\label{tab:ionic_sfr_mass_density_classy}
\begin{tabular}{ccccccccc}
\toprule
Galaxy & \begin{tabular}[c]{@{}c@{}}$n_{\rm e}$ [S~{\sc ii}] \\ (cm$^{-3}$)\end{tabular}& \begin{tabular}[c]{@{}c@{}}$n_{\rm e}$ [Ar~{\sc iv}]$\dagger$ \\ (cm$^{-3}$)\end{tabular} & \begin{tabular}[c]{@{}c@{}}12+ \\ log(O/H)\end{tabular} & \begin{tabular}[c]{@{}c@{}}log(N/O) \\ \end{tabular} & \begin{tabular}[c]{@{}c@{}}12+ \\ log(N/H)\end{tabular} & \begin{tabular}[c]{@{}c@{}}log($\Sigma_{\mathrm{SFR}}$) \\ (M$_\odot$ yr$^{-1}$ kpc$^{-2}$)\end{tabular} & \begin{tabular}[c]{@{}c@{}}log($\Sigma_{\star}$) \\ (M$_\odot$ kpc$^{-2}$)\end{tabular} & \begin{tabular}[c]{@{}c@{}}WR\\Ref.\end{tabular}  \\
\midrule
J0021+0052 & $<100$       & - & $8.15 \pm 0.09$ & $-1.20 \pm 0.23$ & $6.95 \pm 0.14$ & $-0.04^{+0.14}_{-0.11}$ & $1.98^{+0.18}_{-0.38}$ & - \\
J0036-3333 & $<100$       & -&  $8.16 \pm 0.06$ & $-0.96 \pm 0.11$ & $7.20 \pm 0.09$ & $0.06^{+0.19}_{-0.21}$ & $2.19^{+0.26}_{-0.23}$ & - \\
J0127-0619 & $408 \pm 40$ & - & $8.13 \pm 0.05$ & $-0.88 \pm 0.23$ & $7.25 \pm 0.13$ & $-0.39^{+0.15}_{-0.13}$ & $3.10^{+0.18}_{-0.15}$ & 1 \\
J0144+0453 & $<100$       & - &  $7.45 \pm 0.08$ & $-1.61 \pm 0.25$ & $5.83 \pm 0.22$ & $-0.58^{+0.29}_{-0.46}$ & $1.88^{+0.24}_{-0.29}$ & - \\
J0337-0502 & $180 \pm 10$ & $930\pm150$ & $7.23 \pm 0.01$ & $-1.32 \pm 0.08$ & $5.91 \pm 0.06$ & $-0.31^{+0.07}_{-0.11}$ & $1.07^{+0.24}_{-0.21}$ & - \\
J0405-3648 & $<100$       & - & $7.28 \pm 0.07$ & $-1.30 \pm 0.19$ & $5.98 \pm 0.15$ & $-1.24^{+0.31}_{-0.27}$ & $1.18^{+0.28}_{-0.28}$ & - \\
J0808+3948 & $1179 \pm 100$  & $6310\pm3600$ & $8.77 \pm 0.05$ & $-0.76 \pm 0.06$ & $7.92 \pm 0.08$ & $-0.09^{+0.18}_{-0.25}$ & $1.77^{+0.30}_{-0.17}$ & - \\
J0823+2806 & $144 \pm 24$    & $5100\pm4300$ & $8.25 \pm 0.10$ & $-1.08 \pm 0.26$ & $7.16 \pm 0.16$ & $0.09^{+0.15}_{-0.32}$ & $1.99^{+0.33}_{-0.19}$ &  2\\
J0926+4427 & $<100$ & - &  $7.97 \pm 0.06$ & $-1.83 \pm 0.19$ & $6.14 \pm 0.14$ & $-0.63^{+0.13}_{-0.13}$ & $1.10^{+0.30}_{-0.26}$ & - \\
J0934+5514 & $100$  & $8240\pm4310$  & $7.09 \pm 0.02$ & - & - & $-0.58^{+0.09}_{-0.07}$ & $1.21^{+0.15}_{-0.20}$ & - \\
J0938+5428 & $106 \pm 37$ & - & $8.26 \pm 0.08$ & $-1.24 \pm 0.22$ & $7.02 \pm 0.15$ & $-0.37^{+0.20}_{-0.17}$ & $1.73^{+0.18}_{-0.29}$ & - \\
J0940+2935 & $<100$& - & $7.98 \pm 0.23$ & $-1.75 \pm 0.36$ & $6.23 \pm 0.20$ & $-1.31^{+0.42}_{-0.37}$ & $1.41^{+0.23}_{-0.40}$ & - \\
J0942+3547 & $<100$ & -& $8.01 \pm 0.07$ & $-1.20 \pm 0.28$ & $6.81 \pm 0.16$ & $-0.78^{+0.19}_{-0.12}$ & $1.54^{+0.21}_{-0.29}$ & - \\
J0944+3442 & $113 \pm 48$ & -& $7.66 \pm 0.15$ & $-0.99 \pm 0.32$ & $6.67 \pm 0.19$ & $-0.35^{+0.19}_{-0.16}$ & $1.26^{+0.44}_{-0.25}$ & - \\
J0944-0038 & $138 \pm 56$ & $1060\pm590$ & $7.83 \pm 0.02$ & $-1.14 \pm 0.19$ & $6.69 \pm 0.11$ & $-0.01^{+0.28}_{-0.65}$ & $2.19^{+0.40}_{-0.23}$ & - \\
J1016+3754 & $<100$ & -& $7.57 \pm 0.03$ & $-1.37 \pm 0.22$ & $6.19 \pm 0.14$ & $-0.30^{+0.18}_{-0.18}$ & $1.59^{+0.27}_{-0.22}$ & - \\
J1024+0524 & $<100$& - & $7.80 \pm 0.05$ & $-1.40 \pm 0.22$ & $6.41 \pm 0.13$ & $-0.47^{+0.14}_{-0.12}$ & $1.21^{+0.37}_{-0.24}$ & - \\
J1025+3622 & $198 \pm 56$& - & $8.13 \pm 0.08$ & $-1.36 \pm 0.23$ & $6.77 \pm 0.14$ & $-0.32^{+0.14}_{-0.18}$ & $1.51^{+0.25}_{-0.27}$ & - \\
J1044+0353 & $267 \pm 19$ & $404\pm110$ & $7.55 \pm 0.02$ & $-1.48 \pm 0.23$ & $6.07 \pm 0.14$ & $-0.39^{+0.11}_{-0.14}$ & $1.00^{+0.41}_{-0.26}$ & - \\
J1105+4444 & $113 \pm 56$ & $3160\pm2600$ & $8.23 \pm 0.07$ & $-1.46 \pm 0.22$ & $6.77 \pm 0.15$ & $-0.23^{+0.28}_{-0.22}$ & $2.06^{+0.29}_{-0.24}$ & - \\
J1112+5503 & $407 \pm 19$& - & $8.02 \pm 0.04$ & $-0.65 \pm 0.07$ & $7.36 \pm 0.06$ & $0.14^{+0.20}_{-0.25}$ & $2.13^{+0.33}_{-0.19}$ & - \\
J1119+5130 & $<100$ & -& $7.59 \pm 0.08$ & $-1.59 \pm 0.25$ & $5.99 \pm 0.16$ & $-0.85^{+0.21}_{-0.12}$ & $1.50^{+0.15}_{-0.28}$ & - \\
J1129+2034 & $<100$& - & $8.30 \pm 0.12$ & $-1.48 \pm 0.32$ & $6.82 \pm 0.19$ & $-0.12^{+0.38}_{-0.56}$ & $2.34^{+0.37}_{-0.27}$ & 2 \\
J1132+1411 & $<100$& - & $8.24 \pm 0.07$ & $-1.45 \pm 0.24$ & $6.79 \pm 0.15$ & $-0.17^{+0.24}_{-0.27}$ & $2.07^{+0.28}_{-0.19}$ & - \\
J1132+5722 & $122 \pm 41$ & -& $7.34 \pm 0.09$ & $-1.43 \pm 0.21$ & $5.91 \pm 0.15$ & $-1.63^{+0.27}_{-0.35}$ & $0.75^{+0.23}_{-0.26}$ & - \\
J1144+4012 & $109 \pm 48$& - & $8.65 \pm 0.08$ & $-1.20 \pm 0.13$ & $7.45 \pm 0.12$ & $-0.13^{+0.20}_{-0.29}$ & $2.25^{+0.18}_{-0.29}$ & - \\
J1148+2546 & $113 \pm 10$ & $1190\pm1070$ & $8.01 \pm 0.04$ & $-1.39 \pm 0.21$ & $6.62 \pm 0.14$ & $-0.02^{+0.17}_{-0.14}$ & $1.59^{+0.34}_{-0.24}$ & - \\
J1150+1501 & $<100$ & $1200\pm920$ & $8.14 \pm 0.10$ & $-1.52 \pm 0.31$ & $6.63 \pm 0.17$ & $-0.03^{+0.29}_{-0.23}$ & $2.14^{+0.28}_{-0.30}$ & -\\
J1157+3220 & $<100$& - & $8.46 \pm 0.22$ & $-1.42 \pm 0.37$ & $7.04 \pm 0.21$ & $0.55^{+0.21}_{-0.42}$ & $2.62^{+0.32}_{-0.18}$ & 3 \\
J1200+1343 & $172 \pm 53$& - & $8.16 \pm 0.06$ & $-1.04 \pm 0.24$ & $7.12 \pm 0.14$ & $-0.18^{+0.20}_{-0.16}$ & $1.19^{+0.47}_{-0.42}$ & - \\
J1225+6109 & $<100$& - & $8.02 \pm 0.05$ & $-1.65 \pm 0.31$ & $6.37 \pm 0.15$ & $-0.08^{+0.26}_{-0.26}$ & $2.12^{+0.34}_{-0.24}$ & 4 \\
J1253-0312 & $437 \pm 35$ & $470\pm320$& $8.02 \pm 0.03$ & $-0.92 \pm 0.21$ & $7.09 \pm 0.12$ & $0.37^{+0.15}_{-0.15}$ & $1.46^{+0.51}_{-0.23}$ & - \\
J1314+3452 & $180 \pm 15$ & - & $8.27 \pm 0.11$ & $-1.52 \pm 0.30$ & $6.76 \pm 0.18$ & $0.49^{+0.23}_{-0.55}$ & $2.72^{+0.30}_{-0.21}$ & 2 \\
J1323-0132 & $628 \pm 100$ &$214\pm140$ & $7.72 \pm 0.01$ & $-1.16 \pm 0.18$ & $6.55 \pm 0.11$ & $-0.52^{+0.08}_{-0.09}$ & $0.51^{+0.26}_{-0.10}$ & - \\
J1359+5726 & $<100$ & $2070\pm 2240$ & $7.99 \pm 0.07$ & $-1.38 \pm 0.22$ & $6.62 \pm 0.14$ & $-0.32^{+0.20}_{-0.14}$ & $1.67^{+0.31}_{-0.26}$ & - \\
J1416+1223 & $295 \pm 61$ &- & $8.15 \pm 0.04$ & $-0.84 \pm 0.06$ & $7.31 \pm 0.06$ & $0.10^{+0.21}_{-0.25}$ & $2.12^{+0.32}_{-0.26}$ & - \\
J1418+2102 & $<100$ & $1150\pm150$ & $7.51 \pm 0.03$ & $-1.46 \pm 0.23$ & $6.05 \pm 0.15$ & $-0.52^{+0.15}_{-0.16}$ & $0.83^{+0.49}_{-0.35}$ & - \\
J1428+1653 & $119 \pm 62$ &- & $8.21 \pm 0.14$ & $-1.13 \pm 0.25$ & $7.08 \pm 0.17$ & $-0.49^{+0.26}_{-0.19}$ & $1.85^{+0.15}_{-0.23}$ & - \\
J1429+0643 & $<100$ &-& $7.95 \pm 0.06$ & $-1.11 \pm 0.19$ & $6.84 \pm 0.14$ & $-0.18^{+0.11}_{-0.17}$ & $1.20^{+0.35}_{-0.21}$ & - \\
J1444+4237 & $<100$&- & $7.38 \pm 0.09$ & $-1.20 \pm 0.30$ & $6.19 \pm 0.20$ & $-0.98^{+0.11}_{-0.08}$ & $1.44^{+0.17}_{-0.17}$ & - \\
J1448-0110 & $120 \pm 29$ &-& $8.07 \pm 0.03$ & $-1.36 \pm 0.21$ & $6.71 \pm 0.14$ & $0.05^{+0.13}_{-0.14}$ & $1.27^{+0.41}_{-0.24}$ & 2\\
J1521+0759 & $<100$ &- & $8.67 \pm 0.07$ & $-1.43 \pm 0.16$ & $7.24 \pm 0.11$ & $-0.32^{+0.16}_{-0.17}$ & $1.73^{+0.29}_{-0.30}$ & - \\
J1525+0757 & $186 \pm 60$ &-& $8.76 \pm 0.05$ & $-1.22 \pm 0.08$ & $7.53 \pm 0.08$ & $-0.35^{+0.69}_{-0.24}$ & $2.75^{+0.28}_{-0.42}$ & - \\
J1545+0858 & $153 \pm 15$ &$1750\pm260$ & $7.75 \pm 0.02$ & $-1.40 \pm 0.18$ & $6.35 \pm 0.13$ & $-0.24^{+0.13}_{-0.17}$ & $0.91^{+0.43}_{-0.26}$ & - \\
J1612+0817 & $484 \pm 88$ &-& $8.47 \pm 0.04$ & $-0.96 \pm 0.07$ & $7.51 \pm 0.07$ & $0.06^{+0.28}_{-0.24}$ & $2.26^{+0.28}_{-0.26}$ & - \\
\bottomrule
\end{tabular}
\begin{description}
\item [References] (1) \cite{james09}, (2) WR candidates with FWHM $\ga 500$ km/s (see Sec.~\ref{sec.NO-OH}), (3) \cite{Miralles-caballero16}, (4) \cite{Paswan18}. $\dagger$ The \Ne[\ion{Ar}{iv}] derived in this study are in agreement with the values reported in \citet{mingozzi22}.
\end{description}
\end{table*}
 %


\section{Results}\label{results} \label{sec.NO-OH}
\subsection{The N/O versus O/H relation} \label{sec:label:NO-scatter}
\begin{figure*}
\begin{center}
    \includegraphics[width=0.8\textwidth, trim=30 0 30 0,  clip=yes]{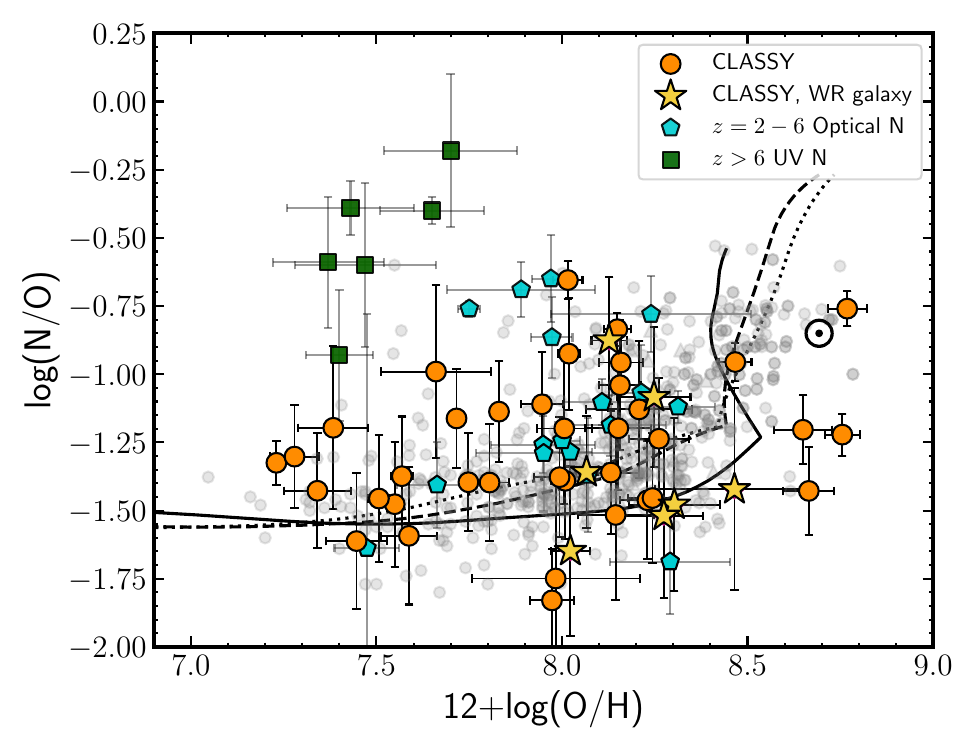}
        \caption{The N/O-O/H relation for the CLASSY sample. CLASSY galaxies hosting WR stars are also identified (star symbols), and they show similar N/O ratios to the bulk of the sample, with the exception of the WR galaxy Mrk~996 \citep{james09}, which exhibits a N/O enhancement. The comparison sample of local SFGs and \ion{H}{ii} regions (grey symbols) was compiled and re-analyzed in \citet{arellanocordova24}. A sample of high-redshift galaxies is included for comparison. SFGs at $z = 2-6$ with N/O ratios derived from optical [\ion{N}{ii}] lines are shown as pentagons \citep[][]{sanders23, welch25, rogers24, arellanocordova25, scholte25, Zhang25}, while squares represent $z > 6$ galaxies with N/O derived from UV nitrogen lines \citep[][]{sanders23b, marqueschaves24, jones23, arellanocordova24, schaerer24, isobe23a, nakajima23, topping24b}. 
        The different curves show predictions from chemical evolution models with an infall mass of log($M_{\rm inf}) = 10 $ M$_\odot$, star formation efficiencies (SFE) of 0.5, 1, and 5 Gyr$^{-1}$, and a time infall of $\tau_{\rm inf} = 0.1$ Gyr represented by solid, dashed, and dotted lines, respectively \citep{vincenzo16}.
        The CLASSY galaxies (orange circles) follow the expected N/O-O/H relation and occupy a similar parameter space as SFGs at $z=2-6$. In contrast, they differ from the $z>6$ population, showing no apparent redshift evolution.}
\label{fig:NO_highz}
\end{center}
\end{figure*}

In Fig.~\ref{fig:NO_highz}, we present the N/O ratio as a function of metallicity for CLASSY. Overall, we show that CLASSY galaxies follow a constant behavior between N/O and metallicity. On the other hand, the CLASSY galaxies follow the plateau defined by dwarf SFGs (gray symbols) due to the primary contribution of nitrogen attributed mainly to massive stars. For higher metallicities, the CLASSY sample also shows high values of N/O at a fixed metallicity, consistent with the sample compiled from the literature from dwarf galaxies and \hii\ regions \citep[see also][]{arellanocordova24}. In particular, at 12+log(O/H) $\sim$ 8, CLASSY shows a large variation of the log(N/O) of up to 1.5 dex. 
We additionally highlight the seven galaxies exhibiting WR features with yellow stars. Four of them show broad \ion{He}{ii}~\W4686 emission lines in the optical CLASSY spectra, with full width at half maximum (FWHM) $\ga$ 500 km/s. The other three have WR features confirmed in previous studies \citep[e.g.,][]{james09, Miralles-caballero16, Paswan18} (see Table~\ref{tab:ionic_sfr_mass_density_classy}). As shown in Fig.~\ref{fig:NO_highz}, the CLASSY galaxies with WR features tend to be located at high metallicities (12 + log(O/H) $>$ 8.0) and low log(N/O) values ($<-1.30$) following the bulk of the observations, with the exception of J0127-0619 and J0823+2806, which shows slightly high  N/O ratio than the rest of the sample hosting WR stars \citep[Mrk~996;][]{james09}. However, note that these WR galaxies are also consistent with the N/O ratios of local SFGs and \hii\ regions for a fixed metallicity. In Fig.~\ref{fig:NO_highz}, we have incorporated a sample of high-redshift galaxies in comparison with CLASSY. 
 Starting from the optical N/O results, we can see that the CLASSY galaxies cover the parameter spaces of $z=2-6$ galaxies. While at $z > 6$, the sample of SFGs shows a significantly high N/O for 12+log(O/H) $< 8$ as reported in their original studies. Moreover, the  UV and optical N/O abundances occupy two different positions in the N/O-O/H relation. This could suggest that optical and UV lines trace different zones of the nebula \citep[e.g.,][]{topping24, pascale23}. In Sec.~\ref{sec:discussion}, we will provide a detail comparison of CLASSY and high-$z$ galaxies. 

 To inspect the chemical enrichment of the CLASSY galaxies in the N/O-O/H relation, we have included chemical evolutions models of \citet{vincenzo16} in Fig.~\ref{fig:NO_highz} \citep[see also][]{kumari18, hayden-pawson22}. We have selected three different models covering the following parameters: an infall mass of log(M$_{\rm infall}$/M$_{\odot}$) = 10, star formation efficiency (SFE) of $\nu=$  0.5 Gy$^{-1}$ (dotted), $\nu=$ 1 Gy$^{-1}$ (dashed), and $\nu=$ 5 Gy$^{-1}$ (solid), time fall $\tau= 1$ Gyr and a mass loading factor $\omega = 1.0$. These models assume a \citet{salpeter55} IMF. 
Overall, most CLASSY galaxies follow the evolutionary track by the chemical evolution models, but the observations show high scatter.  
However, most galaxies exhibit higher N/O ratios at a fixed metallicity. In principle, a higher SFE should result in an increase in N/O due to the early contribution of intermediate massive stars \citep{vincenzo16}.  
We stress that, since galaxies have different star-formation histories, it is necessary to compare the observed N/O–O/H relation with models different star formation history and feedback in order to properly interpret the observed relation in Fig.~\ref{fig:NO_highz} \citep[see e.g.,][]{kobayashi24, Bhattacharya25}.

\subsection{The N/H versus O/H relation}
\begin{figure}
\begin{center}

    \includegraphics[width=0.42\textwidth, trim=30 0 30 0,  clip=yes]{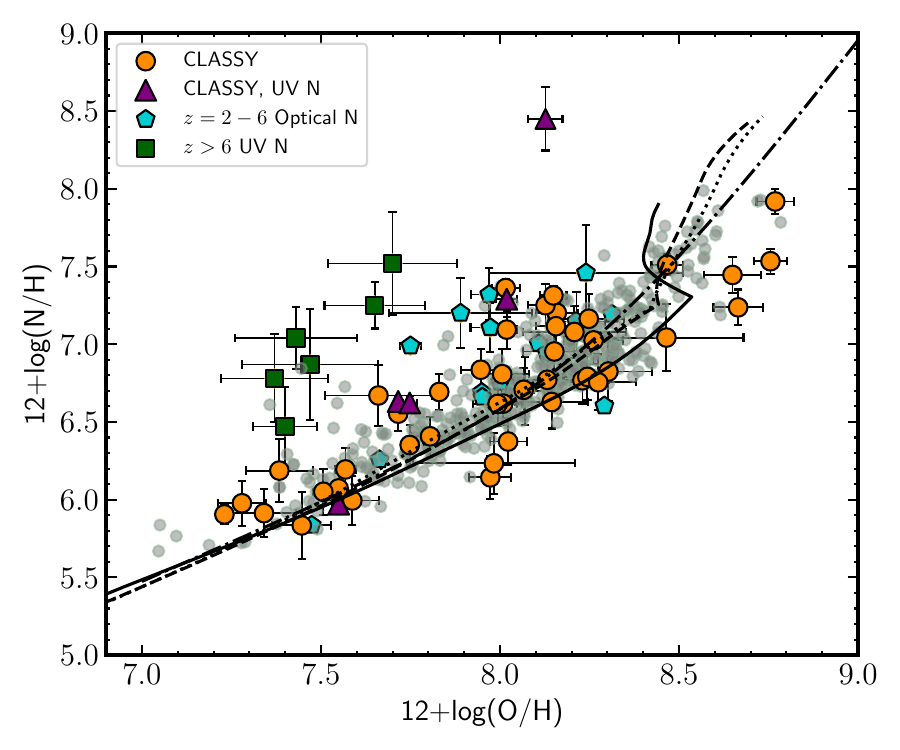}
        \caption{The N/H-O/H relation for CLASSY and the sample of high redshift galaxies. SFGs at $z=2-6$ from \citet{sanders23b, welch25, rogers24, arellanocordova25, scholte25} and \citet{Zhang25} in cyan pentagons, while the green squares represent $z>6$ galaxies from \citet{marqueschaves24, topping24, topping24b, curti24b, schaerer24}. For comparison, we include the local sample in gray circles. High redshift galaxies are shifted to higher values of N/H, mainly those galaxies using UV nitrogen lines (squares), while the optical N sample at $z =2-6$ follows the trend of CLASSY at $z\sim0$. The purple triangles show those CLASSY galaxies with N/O derived using UV [\ion{N}{iv}] or [\ion{N}{iii}] lines (see Martinez et al. 2025).  
        The different curves show the predictions of chemical evolution models for the N/H–O/H relation, as in Fig.~\ref{fig:NO_highz}, while the dash-dotted line represents an empirical calibration to SDSS galaxies from \citet{flury20}.}
\label{fig:NH_highz}
\end{center}
\end{figure}

We also explore the relation between the total N/H and O/H for CLASSY shown in Fig.~\ref{fig:NH_highz}. Additionally, we have added the sample of local SFGs and \ion{H}{ii} regions from the literature. As expected, we find a correlation between these two abundance ratios due to nucleosynthesis of intermediate- and massive stars. For comparison, we have added the same chemical evolution models as in Fig.~\ref{fig:NO_highz}, together with the empirical N/H–O/H relation for SDSS SFGs from \citet[][]{flury20}, which is consistent with the observed trends.

Then, we have also incorporated the high-redshift for comparison in  Fig.~\ref{fig:NH_highz}. First, we found that galaxies with optical N abundance determination follow the same abundance pattern as CLASSY for a metallicity range of 12+log(O/H) = 7.5-8.5. In contrast, galaxies at $z>6$ showing 12+log(O/H) $<$ 8.0 are shifted to higher values of 12+log(N/H). This might be due to the high N/O and very low metallicity of $z>6$ galaxies
(green squares) derived from their physical conditions \citep[][Martinez et al. 2025]{hayes25}.
As an inspection, we have added the N/H values for five galaxies with UV N emission lines (J0127-0639, J1253-0132, J1044+0353, and J1545+058) reported in Martinez et al. 2025. Fig.~\ref{fig:NH_highz} shows that CLASSY galaxies with N/H derived using both UV (purple triangles) and optical N lines exhibit a similar abundance patterns within the uncertainties. Overall, the differences between UV and optical N/O abundances for CLASSY can reach 0.08-0.28 dex in N/H, with the exception of J0127-0639, which exhibits a difference of 1.2 dex. 
 J0127-0639 (or Mrk996) is an object with complex kinematics and chemical enrichment. As seen in Fig~\ref{fig:NH_highz}, Mrk~996 exhibits a very high N/H ratio that departs significantly from the bulk of the sample. This difference is likely due to the presence of WR stars in combination with their exotic kinematics, which can enrich N/O regardless of whether optical or UV nitrogen lines are used \citep[][]{james09} (see also Fig.~\ref{fig:NO_highz}). A detailed discussion of UV N/O galaxies (including Mrk~996) is presented in Martinez et al. 2025.
 In the following sections, we will analyze the physical conditions and properties of the CLASSY galaxies to better understand the high scatter in the N/O-O/H relation.

\subsection{Electron Density}\label{sec:density}
The electron density, usually calculated using the [\ion{S}{ii}]~\W\W6717,31 lines, and \Te\ are crucial physical conditions to determine the chemical composition of the gas. For objects with \Ne\ $\leq$100 cm$^{-3}$, the inferred abundances are less sensitive to density \citep[e.g.,][]{osterbrock89, peimbert17}. For high density objects \Ne $\geq$300 cm$^{-3}$, the N/O is less explored since if density diagnostics are not available or are very uncertain might introduce bias in the determination of N/O \citep[e.g.,][]{stasinka23, Zhang25}. \citet{mingozzi22} presented an analysis of the density structure of CLASSY galaxies by comparing density estimates that correspond to low-, intermediate, and high-ionization regions (such as \Ne[\ion{S}{ii}], \Ne[\ion{Cl}{iii}], and \Ne[\ion{Ar}{iv}], respectively). These authors found no significant impact on metallicity using  different density diagnostics on the metallicity inferred for the CLASSY galaxies, while N/O was not explored in that study. 

\begin{figure*}
\begin{center}
    \includegraphics[width=0.9\textwidth, trim=10 0 10 0,  clip=yes]{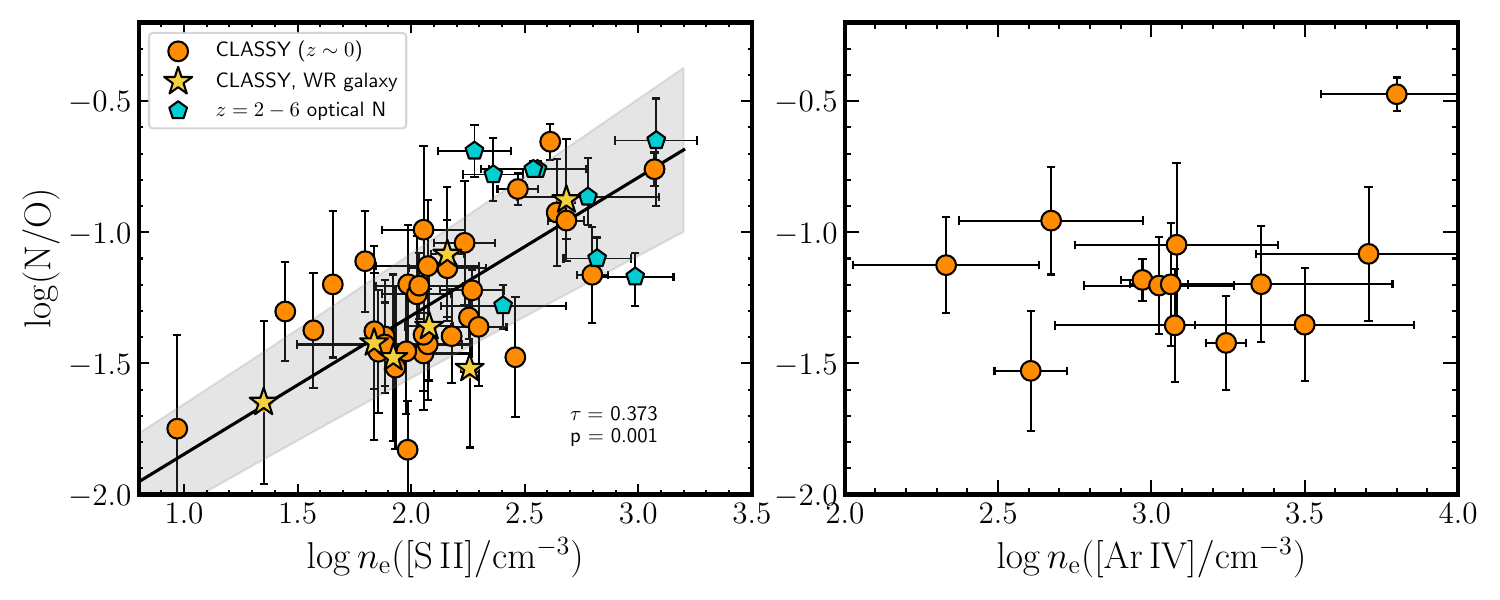}

        \caption{ \textit{Left} N/O abundances as a function of low-ionization density, \Ne[\ion{S}{ii}]. SFGs at  $z>2$ are also included for comparison from \citep{sanders23, welch25, arellanocordova25, stanton2025, scholte25}. We find that high density derived from $n_{\rm e}$[\ion{S}{ii}] correlates with the high values of N/O for local galaxies (with a Kendall's coefficient of $\tau=0.372$ and p-value = 0.001), while $z=2-6$ galaxies show a similar trend. The yellow stars are galaxies with WR stars features. The best fit to the data for the N/O {\it vs.} log($n_{\rm e}$[\ion{S}{ii}]) is shown in an orange solid line for CLASSY. \textit{Right}: The N/O ratios as a function of high-ionization density, \Ne[\ion{Ar}{iv}]. N$^{+}$ and O$^{+}$ were derived using \Ne[\ion{Ar}{iv}] for 13 CLASSY galaxies. Only three CLASSY galaxies have measurements of both \Ne[\ion{S}{ii}] and \Ne[\ion{Ar}{iv}] within a similar range of \Ne\ $=$ 200–600 cm$^{-3}$. J0808+3948 shows the highest values, with \Ne\ $=$ 1200–6310 cm$^{-3}$ and log(N/O) = –0.47 to –0.65 when derived using both density diagnostics.}
\label{fig:density-structure}
\end{center}
\end{figure*}

 In the left panel of Fig.~\ref{fig:density-structure}, we compare the N/O ratio as a function of \Ne [\ion{S}{ii}]. We have used the Kendall's $\tau$ coefficient to verify any potential correlation between N/O and \Ne [\ion{S}{ii}]. We find a correlation between N/O and \Ne [\ion{S}{ii}] (with a p-value = 0.001,  $\tau = 0.373$), showing that high density objects, log(\Ne[\ion{S}{ii}]/cm$^{-3}$) $\geq 2.5$ are shifted to high N/O ratios.

 Our best fit  using orthogonal distance regression \citep[ODR,][]{boggs81, scipy} to the relation between N/O and \Ne\ is as follow:

 \begin{equation}
\mathrm{log(N/O)} = (0.432\pm0.071)\times \mathrm{log(n_{\rm e})} - (2.166\pm0.0150).
 \end{equation}

To analyze the result presented in the left of Fig.~\ref{fig:density-structure}, we have also determined the N/O ratios using \Ne[\ion{Ar}{iv}], which traces the high ionization density structure of the gas. For 13 CLASSY galaxies we recalculated the N/O ratios adopting \Ne[\ion{Ar}{iv}] instead of \Ne[\ion{S}{ii}] (see the right panel of Fig.~\ref{fig:density-structure}). For such galaxies, we measure 2.5 < log(\Ne[\ion{Ar}{iv}]/cm$^{-3}$) < 3.2, while the same galaxies also show 2 < log(\Ne[\ion{S}{ii}]/cm$^{-3}$) < 3. Note that the density range derived for these galaxies depends on the sensitivity of the diagnostic used \citep[e.g.,][]{mendez-delgado+23a}. We find that the differences in N/O using different density measurements can reach values of up to 0.2-0.3 dex higher, with some exception showing values up to 0.1 dex lower. It implies that N/O might be affected by the density structure of the gas, which is particularly associated with the [\ion{O}{ii}]~\W\W3726,29 to calculate O$^{+}$ \citep[see][]{mendez-delgado23b, stasinka23}, due to the low critical density of those lines \citep{juandedios16}. In contrast to the correlation between N/O and \Ne[\ion{S}{ii}], there is no correlation between N/O and \Ne[\ion{Ar}{iv}] (see the right panel of Fig.~\ref{fig:density-structure}). Of the seven CLASSY galaxies with high N/O and \Ne[\ion{S}{ii}], only three have \Ne[\ion{Ar}{iv}] measurements (see Table \ref{tab:ionic_sfr_mass_density_classy}). While two of them show good agreement between the values of \Ne[\ion{S}{ii}] and \Ne[\ion{Ar}{iv}], J0808+3948, with high density in both diagnostics, shows an increase in log(N/O), with a value 0.28 dex higher when \Ne[\ion{Ar}{iv}] is used. A larger sample of galaxies with both \Ne[\ion{S}{ii}] and \Ne[\ion{Ar}{iv}] is needed to better understand the role of \Ne\ in the high N/O ratios of galaxies at \Ne$ > 300$ cm$^{-3}$ \citep[e.g.,][Martinez et al. 2025]{mendez-delgado+23a, yanagisawa24b}.

On the other hand, in \citet{peng25}, the authors determined the gas density for 14 CLASSY galaxies, reporting measurements of [\ion{O}{ii}]~\W\W3726,29 and [\ion{S}{ii}]~\W\W6717,31 for different gas components derived from the high-resolution spectra. Interestingly, \Ne[\ion{S}{ii}]\W\W6717,31 derived from the narrow component is relatively higher than the values analyzed in this study \citep[e.g.,][see also AC24a]{arellanocordova22a}. For example, for J0808+3948, the object with the highest \Ne[\ion{S}{ii}]~\W\W6717,31 in the CLASSY sample, we derived a value of \Ne[\ion{S}{ii}] $\sim$ 1200 cm$^{-3}$, while the value reported in \citet{peng25} is a factor of 3.5 higher. In contrast, their values for \Ne[\ion{O}{ii}] are aligned with our measurements. We give a look at the values reported for \Ne[\ion{S}{ii}] for the 14 CLASSY galaxies reported in \citet{peng25} from the narrow component, and we find significant differences of up to 650~cm$^{-3}$ higher for most of the galaxies than our measurements. We think that aperture effects should be minimal since in \citet{arellanocordova22a}, these authors show a good agreement between \Ne[\ion{S}{ii}] derived in galaxies with different aperture sizes and instruments when the relative fluxes are used to derive the physical conditions and metallicities. A follow-up of the density structure of CLASSY will provide information on its impact on the chemical abundance ratios in the high density regime. Since \citet{peng25} did not report the N/O ratios in their study, we have derived N/O using the extinction-corrected fluxes of the narrow component and the physical conditions of their study. In this inspection to the N/O ratios, we find consistent results with our result in Fig.~\ref{fig:density-structure}, implying that N/O increases as \Ne\ increases for most of the galaxies of the sample. 
In Fig.~\ref{fig:density-structure}, we have also identified those CLASSY galaxies with WR star features, which are located in the low density regime (log(\Ne[\ion{S}{ii}]/cm$^{-3}$) $ \sim$2, with exception of Mrk~996).
Moreover, for comparison, we added the high-$z$ sample with optical N/O ratios. We find that high density galaxies at $z = 2-6$ with high N/O aligns with our finding at $z\sim0$.

Therefore, from Fig.~\ref{fig:density-structure}, we suggest that part of the dispersion in the N/O-O/H relation can be associated with the density structure of the nebula, implying that intrinsically high density objects (\Ne$\ga 300 $ cm$^{-3}$) might result in high N/O abundances. However, a large sample of galaxies covering high density structure by different diagnostics is necessary to confirm the results present in Fig.~\ref{fig:density-structure}. In Sec.~\ref{sec:discussion}, we discuss these results in comparison with other studies.

\subsection{Global galaxy properties}\label{sec:Galaxy_properties}
The scatter of the N/O-O/H relation could be related to
different processes such as the star-formation history of the galaxies, the time delay of the primary and secondary contribution of N, the enrichment by WR stars, accretion of metal-poor gas, the temperature and density structure used to determine N/O, the impact of the star-formation efficiency in regulating the yield production of O and N, among others \citep[e.g.,][]{henry00, koppen&hensler05, nava06, amorin10, amorin12, perezmontero13, vincenzo16, berg20, arellano-cordova20, hayden-pawson22}
We investigate the impact of various galaxy properties such as stellar mass, SFR and SFR surface density on the N/O and N/H ratios for CLASSY and its evolution across redshifts. These physical measurements were derived from previous CLASSY papers (see Sec.~\ref{sec: sample}). In our analysis, we include the sample local SFGs and high-$z$ galaxies described in Sec.~\ref{sec:Archival_data}.

Fig.~\ref{fig:NO_scatter_properties} shows the comparison of the N/O and N/H ratios as functions of stellar mass (left), SFR (middle) and specific SFR (sSFR, right). We discuss our findings for the CLASSY sample and their comparison with the properties of galaxies at high-$z$ in Sec.~\ref{sec:discussion}. Note that in Fig.~\ref{fig:NO_scatter_properties}, we have also distinguished galaxies whose N/O is derived either using UV or optical emission lines of N for comparison purposes.

\subsubsection{Stellar mass}\label{sec:Mass_CLASSY}
Stellar mass is strongly linked with the chemical evolution of galaxies. Previous studies have found a positive correlation between N/O and M$_\star$ in galaxies, implying that more massive galaxies are chemically enriched from different generations of stars.\citep[e.g.,][]{perez-montero09, hayden-pawson22, amorin10, strom22}. 

The left panels of Fig.~\ref{fig:NO_scatter_properties} show the relation between stellar mass and N/O (top) and N/H (bottom) for CLASSY. We find a significant positive correlation (Kendall's coefficient of $\tau=$0.293 and $p$-value = 0.005) between N/O and the stellar mass with some scatter across masses. We also note that the sample of dwarf galaxies ($z\sim$0) compiled from the literature  are also in agreement with the distribution of CLASSY (grey circles, see also Sec.\ref{sec:Archival_data}). 

We have used the expression reported in \citet[][see their Eq.~4]{hayden-pawson22} to provide a relation between stellar mass and the N/O ratios calculated using the \Te-sensitive method. The reason for the selection of this relation is due to the shape of the N/O-M$\star$ relation as stellar mass increases, log(M$\star/\rm M_{\odot}) > 10$ \citep[e.g.,][]{perez-montero09, perezmontero13, sanders21}.  
The parameter of the fit are as follows: log(N/O)$_{0}$ = $-1.37\pm0.05$, log(N/O)$_{1}$ = $-0.98\pm0.11$, log(M$_{0}$/M$_{\odot}$) = $9.10\pm0.18$, and $k = 3$, where $k$ is the transition between low (N/O$_{0}$) and high plateaus (N/O$_{1}$) and M$_{0}$ indicates where the shape of the relation start to decrease \citep[][see their Eq.~4]{hayden-pawson22}. From our best fit in Fig~\ref{fig:NO_scatter_properties}, we have found a well-defined plateau of N/O below log(M$\star/\rm M_{\odot}) \sim 9$. At log(M$\star/\rm M_{\odot}) > 9$ the fit rapid increases with N/O. 

In Fig.~\ref{fig:NO_scatter_properties}, we have added the N/O versus stellar mass relation derived by \citet{andrews13} for stacked SDSS galaxies at $z \sim 0$ (solid line), where N/O and O/H were obtained using the \Te-sensitive method. This relation exhibits a break, with two distinct components across different metallicity regimes. At low metallicity (log(M$\star/\rm M_{\odot}) < 8.5$), N/O remains roughly constant with stellar mass, consistent with the behavior seen in the CLASSY sample. In contrast, at higher metallicities, a positive correlation between N/O and M$\star$ is observed. The slope of the high-metallicity N/O-M$\star$ relation slightly overlaps with the CLASSY sample, which shows an offset toward lower stellar masses and higher N/O.
We also compare our results with the relation presented in \citet{hayden-pawson22} (dashed line in Fig~\ref{fig:NO_scatter_properties}), which is  based on a sample of local SFGs than span range of stellar masses (10$^8$-10$^{11}$ $M_{\rm \odot}$) and metallicities derived using strong-line methods. We noticed that for 8 < log(M$_{\star}$/M$_{\odot}$) < 9 this relation slightly aligns with the results of CLASSY. For higher stellar masses, CLASSY shows an elevated N/O ratio with some scatter (see also Sec.~\ref{sec:discussion} for a discussion). 

Additionally, we add the sample  at $z=2-6$ in Fig.~\ref{fig:NO_scatter_properties} as a comparison. The $z=2-6$ galaxies are shifted to higher N/O values compared to the relationship for the local SDSS sample by \citet{hayden-pawson22}.
In Fig.~\ref{fig:NO_scatter_properties}, we  also include the fit derived on the N/O-M$\star$ relation at $z=2-3$ galaxies by \citet[][dashed-dotted line]{strom17}.  The N/O and O/H values were derived using a custom strong-line calibrator, showing a scatter of 0.17 dex around the N/O-M$\star$ relation. The stellar masses for this relation range from $10^{9}$ M$_{\odot}$ to $10^{11}$ M$_{\odot}$. \citet{strom17} found a moderate correlation with the stellar mass. However, there is a considerable scatter in the data of 0.33 dex, mainly attributed to the different timescales for N being ejected into the ISM. In this comparison, we find that CLASSY galaxies are shifted to higher N/O values than the relation for $z=2-3$ galaxies, indicating a steeper relation between N/O and stellar mass. We also note that the sample of $z = 2-6$ galaxies (cyan symbols) shows N/O values that are offset from the relation found by \citet{strom17} at similar redshifts. 

For N/H against M$\star$, we found a positive strong correlation ($\tau=$0.627, $p$-value $<10^{-3}$) as stellar mass increases, probably because the dependency between N/H and O/H (see Fig.~\ref{fig:NH_highz}), which is implicit in the determination of N/H = N/O $\times$ O/H. Our best fit to the data show slope and intercept of $0.391\pm0.043$ and $3.55\pm0.356$ (see the bottom panel of Fig~\ref{fig:NO_scatter_properties}). In the same panel, we also show the relation between N/H and stellar mass from \citet{strom17}, which is flatter than the one calculated in this work for the same range of stellar masses. We also note that the galaxies at high-$z$ are located in the same parameter space than the CLASSY sample at $z\sim0$. 
\citet{strom17} also provided the N/H-M$\star$ relation which is displayed in the bottom panel of Fig~\ref{fig:NO_scatter_properties}. We find that for log(M$_{\star}$/M$_{\odot}$) > 9, CLASSY and the high-$z$ sample show a steeper relation with less scatter in comparison to the relation at $z=2-3$  by \citet{strom17}. 

In summary, we have confirmed that N/O increases with the stellar mass with some of scatter at log(M$_{\star}$/M$_{\odot}$) > 8 which might be associated with the secondary production of N by intermediate massive stars \citep[see also][]{perezmontero13} (see also \ref{sec:discussion}). From our comparison with the high-$z$ sample, there is no apparent redshift evolution up to $z\sim6$.

\subsubsection{SFR and sSFR}\label{sec:SFR_CLASSY}
The relation between N/O and SFR in concert with M$_{\star}$ has previously been studied by \citet{perez-montero09, perezmontero13, hayden-pawson22}.  \citet{perezmontero13} reported any secondary dependence of the N/O-M$_{\star}$ with SFR, which makes N/O insensitive to the inflows of metal-poor gas, while that the opposite was found by \citet{hayden-pawson22} in local galaxies. Such a discrepancy is mainly associated for the different strong-line method selected in each study. 

\begin{figure*}
\begin{center}
    \includegraphics[width=0.92\textwidth, trim=30 0 30 0,  clip=yes]{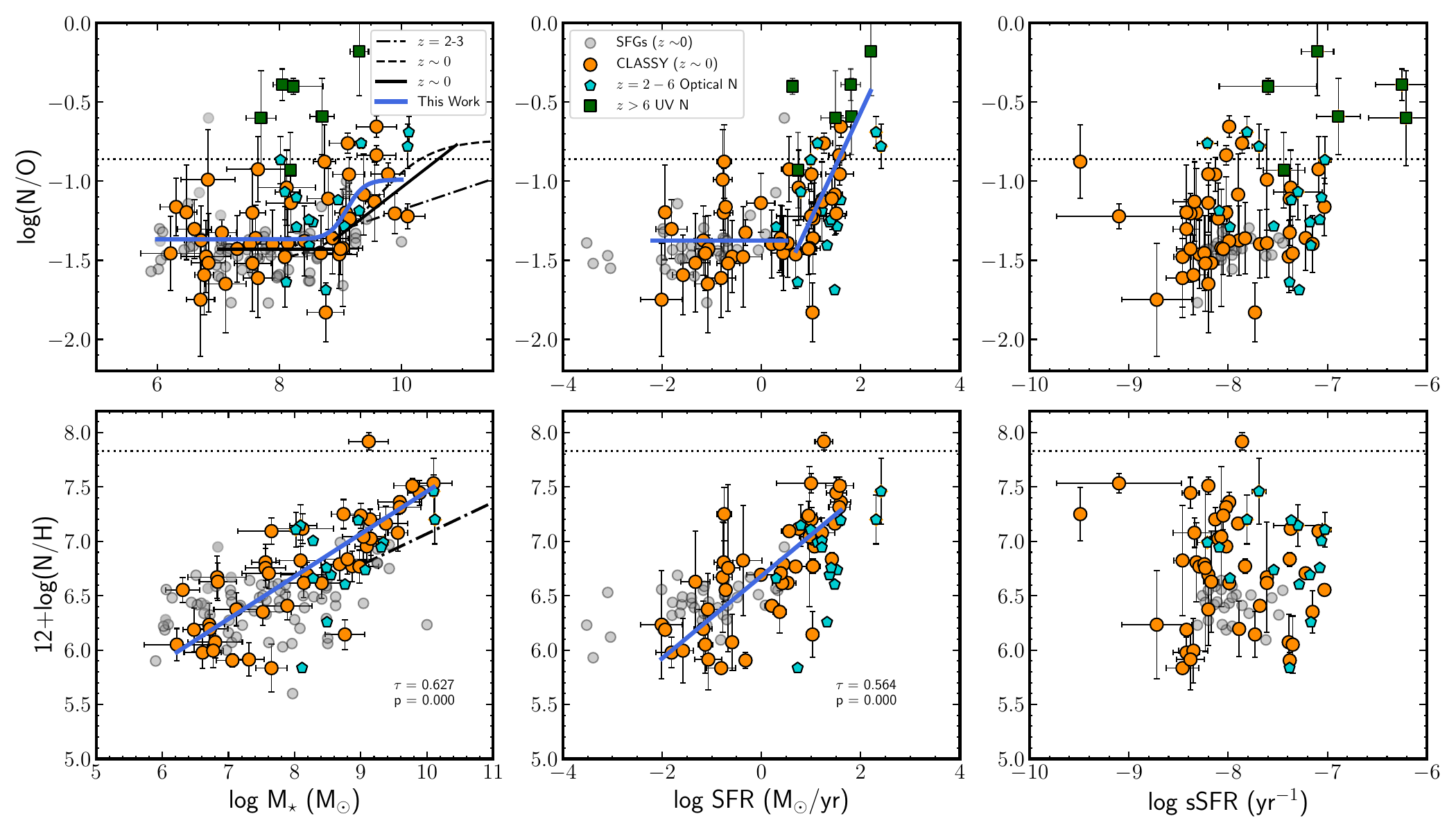}
        \caption{Comparison of the N/O and N/H ratios as a function of the stellar mass (left), SFR (middle) and sSFR (right) for CLASSY. The circles indicate the sample of SFGs from \citet{berg12, berg16, berg19,izotov17}. The dashed line shows the N/O vs M$\star$ for SDSS galaxies derived by \citet{hayden-pawson22}, while the dashed-dotted line represents the N/O \textit{vs.} M$\star$ and  N/H \textit{vs.} M$\star$ relations from \citet{strom22} for galaxies at $z \sim 2-3$. Such relations were derived using strong-line methods. The dotted lines represents the solar value of N/O and N/H from \citet{asplund21}. The pentagons and squares represent galaxies at $z =2-6$ and their N/O abundances on optical lines by \citet{sanders23, welch25, arellanocordova25, scholte25, Zhang25}, while that the squares represent galaxies at $z >6 $ based their N/O abundances on UV lines by \citet{isobe23a, marqueschaves24, topping24, topping24b, curti24b, schaerer24}. Our best fit to the CLASSY data is shown in a solid blue line.}
\label{fig:NO_scatter_properties}
\end{center}
\end{figure*}

Following our analysis with CLASSY with \Te-sensitive chemical abundances, we report the N/O-SFR relation in the top/middle panel of  Fig.~\ref{fig:NO_scatter_properties}.  
The N/O-SFR relation shows a well-defined plateau below  log(SFR/M$_{\odot}$ yr$^{-1}$) $<1$, while there is also a positive correlation with N/O as SFR increases. 
 We have also fitted the N/O-SFR relation using to different slopes indicated in blue on Fig~\ref{fig:NO_scatter_properties}. Our best fits to the data corroborate the plateau at lower SFR with a mean of log(N/O) = $-1.38\pm0.20$. As the SFR increases, the slope of the fit changes abruptly at log(SFR/M$_{\odot}$ yr$^{-1}$)$\sim 1$. For the N/O-SFR plane, we derive log(N/O) = ($0.67\pm0.20$) $\times$ log(SFR/M$_{\odot}$ yr$^{-1}$) $-$ (1.91$\pm0.23$). For the correlation between high N/O and high SFR, we derive a Kendall's coefficient of $\tau = 0.538$ and $p$-value$=0.010$, indicating that ongoing star formation activity probably contributes to the nitrogen production by intermediate massive stars in more evolve systems.  
We also investigated the relation between N/H and SFR, as shown in the bottom/middle panel of Fig.~\ref{fig:NO_scatter_properties}. For N/H, there is a positive correlation, as in the case of N/O, as SFR increases. Our fit shows a smooth increase of N/H with SFR, unlike the abrupt increase observed between N/O and SFR, possibly due to the dependence of O/H on the total abundance of N/H. For the N/H-SFR relation,  we derived a slope and intercept as follow: log(N/H) = ($0.378\pm0.050$) $\times$ log(SFR/M$_{\odot}$ yr$^{-1}$) $+$ (6.68$\pm$0.05), with a Kedall's tau coefficient and $p$-value of $\tau = 0.564$ and $p$-value $< 10^{-3}$, respectively.


Finally, the right column of Fig.~\ref{fig:NO_scatter_properties} shows the relationship between N/O and N/H as a function of specific star formation rate (sSFR). Overall, there is no clear correlation between N/O and sSFR, since the variations in N/O can reach up to 0.5 dex for a fixed sSFR.  Similarly, the comparison between N/H and sSFR shows results akin to those for N/O (see the bottom/right panel of Fig.~\ref{fig:NO_scatter_properties}), with no correlation between these two quantities. 
The fact that SFR and mass together produce no correlation may instead indicate that the long-term and short-term chemical enrichment pathways contribute equally to the N/O abundance pattern. While no correlation exists between N/O and sSFR, significant scatter persists in sSFR at a given N/O for which statistical uncertainty cannot account. This dispersion may indicate physical variations in the mass assembly history (N production, stellar mass) and instantaneous enrichment (O production, SFR) which contribute to the observed abundance patterns.

\subsubsection{$\Sigma_{M_{\star}}$ and $\Sigma_{SFR}$}\label{sec:surface_Density_analysis}
 Previous studies have indicated that the N/O enhancement in galaxies at high-$z$ is related to its compactness  \citep[e.g.,][]{schaerer24, marqueschaves24, reddy23a, reddy23b,   topping25}, which correlates with high star-formation surface density ($\Sigma_{SFR}$) and stellar mass density ($\Sigma_{M_{\star}}$). In this context, we have derived $\Sigma_{\rm SFR}$ and $\Sigma_{M_{\star}}$ for CLASSY to compare with the N/O and N/H abundance ratios at $z\sim0$. These physical properties can provide information of the intensity of SFR encapsulated by the half-light radii ($r_{\rm eff}$) in their impact on the observed abundance patterns of N. For CLASSY, we have taken measurements for $r_{\rm eff}$ from \citet{berg22} (see also Sec.~\ref{sec_sub:gal_properties_comp}). Such as sizes range from 0.78\arcsec to 7.99\arcsec\, ($0.11-2.85$ kpc). Then, we determine stellar mass and SFR surface density as $\Sigma_{SFR}$ = SFR/(2$\pi$$r^{2}_{\rm eff}$) and $\Sigma_{M_{\star}}$ = $M_{\star}$/(2$\pi$$r^{2}_{\rm eff}$). To compare our results with galaxies at $z >2$, we have also compiled $\Sigma_{\rm SFR}$ and $\Sigma_{M_{\star}}$ from the original studies. When these measurements are not available, we calculate $\Sigma_{\rm SFR}$ and $\Sigma_{M_{\star}}$ using their report values of SFR and $r_{\rm eff}$ \citep[i.e.,][]{marqueschaves24, schaerer24, curti24b, topping24, topping24b, Zhang25}. 

\begin{figure*}
\begin{center}
    \includegraphics[width=0.8\textwidth, trim=30 0 30 0,  clip=yes]{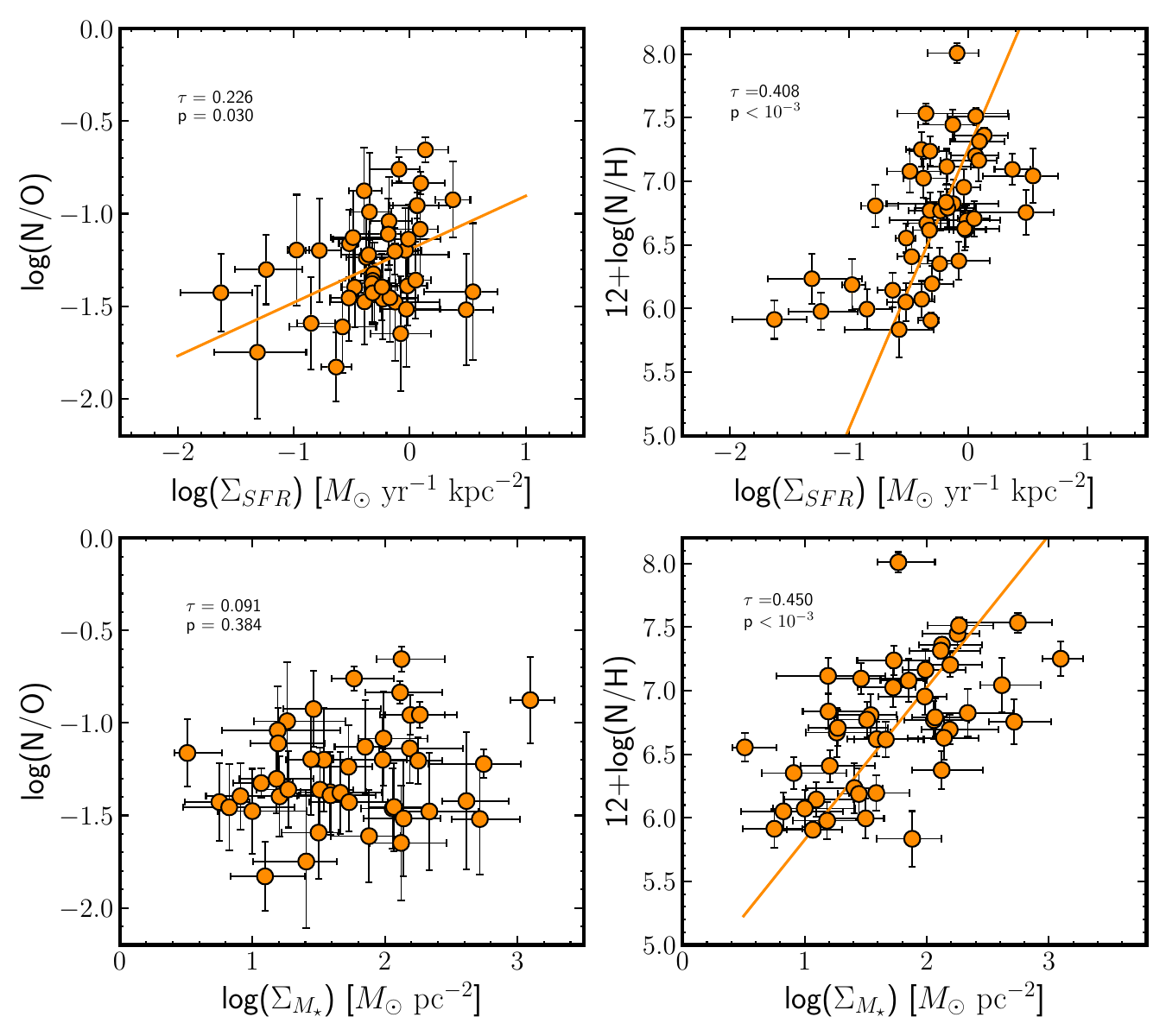}
        \caption{\textit{Top:} The star formation surface density as a function of the N/O and N/H abundance ratios for CLASSY. \textit{Bottom:} Stellar mass surface density as a function of the N/O and N/H abundance ratios for CLASSY. The solid line represents the best fit to the data for CLASSY. The correlation between $\Sigma_{SFR}$ \textit{vs.} N/O indicates that compact objects shows high values of N/O. The Kendall's coefficient of $\tau$ and the $p$-value are indicated in each panel. }
\label{fig:SFR_density}
\end{center}
\end{figure*}

The top panels of Fig.~\ref{fig:SFR_density} shows the comparison of N/O and N/H ratios against $\Sigma_{\rm SFR}$ for CLASSY. We find a moderate positive correlation between N/O and the SFR surface density with Kendall's $\tau = 0.226$ and a $p$-value of 0.030. It implies that galaxies with higher star-formation activity exhibit elevated N/O ratios (see also Fig.~\ref{fig:NO_scatter_properties}), although with some scatter around the fit of $\sigma = 0.25$ dex. We have determined a slope and intercept of $a = 0.29\pm0.09$ and $b=-1.19\pm0.05$, respectively, considering the uncertainties in both axis. For N/H, we report a significant correlation between the 12+log(N/H) and $\Sigma_{\rm SFR}$, with Kendall's coefficient of $\tau=0.408$ and $p$-value of $p$-value $< 10^{3}$. The best fit to the data implies a slope of $a=2.20\pm0.42$ and an intercept of $b=7.25\pm0.16$, respectively, with a scatter around the fit of $\sigma = 0.32$ dex.  

On the other hand, the bottom panel of Fig.~\ref{fig:SFR_density} shows a poor correlation between the N/O \textit{vs.}  $\Sigma_{M_{\star}}$, which is indicated for a Kendall's coefficient of $\tau=0.091$. In contrast, we find a significant correlation between 12+log(N/H) and the stellar mass surface density, probably because of the metallicity is involved in the calculations of N/H (see also Fig.~\ref{fig:NO_scatter_properties}). Our best fit for such relation provides a slope and intercept of $a=1.20\pm0.19$ and $b=4.63\pm0.351$, respectively.

\subsection{Feedback and N/O enrichment}
\label{sec:feedback}
Galactic outflows play an essential role in regulating galaxy evolution \citep[e.g.,][]{heckman11, Heckman15}. Optical emission lines such as H$\beta$ and [\ion{O}{iii}]~$\lambda$5007 have been widely used to trace signatures of outflowing gas associated with winds from massive stars or AGN activity \citep[e.g.,][]{erb15, Chisholm17, mingozzi22, Hogarth20, Flury2023, amorin24, komarova25, peng25}. While optical lines provide valuable information on the kinematics of ionized gas, UV low-ionization absorption lines trace the neutral outflowing gas, offering complementary insights into the multi-phase nature of outflows and the processes that drive them \citep[e.g.,][]{Heckman15, james15, hernandez20, xu22, Flury2023, hayes23, parker24, Xu25}.

To understand the impact of outflows driven by stellar winds on the N/O ratio, we use the offset velocity ($v_{\rm out}$) and mass outflow rate ($\dot{M}_{\rm out}$) measurements reported by \citet{xu22} for CLASSY using UV absorption lines. 
In the bottom panel of Fig.~\ref{fig:outflows}, we show a moderate correlation between outflow velocity and the N/O ratio, color-coded by $v_{\rm out}$ (Kendall's $\tau = -0.472$, $p$-value = 0.002). 
This correlation implies that CLASSY galaxies with high-velocity outflows tend to exhibit elevated N/O ratios (see also James et al. 2025 for a multiphase gas analysis of N). We also note that galaxies with high-velocity outflows generally have relatively high metallicities and high \Ne. These findings suggest that the scatter in the N/O-O/H relation at 12 + log(O/H) $\gtrsim 8$ reflects an interplay between SFR, electron density, and feedback-driven enrichment. Indeed, empirical scaling relations between stellar mass, SFR, metallicity, outflow velocity, and gas content together imply this feedback-driven interplay among these properties \citep[e.g.,][]{Chisholm18, Flury2023}.

Additionally, we calculate the mass-loading factor ($\eta = \dot{M}_{\rm out}/{\rm SFR}$), which indicates the efficiency of feedback (or the relative strength of the outflow) from young (1-10 Myr) stellar populations normalized to the number of massive stars. {There are important implicit dependencies in the estimation of $\dot{M}_{\rm out}$ that should be kept in mind in our analysis of feedback and chemical enrichment, such as $v_{\rm out}$ and the outflow radius \citep[see Eq. 13 in][]{xu22}.  \citet{xu22} reported a strong anti-correlation between the mass-loading factor and both $v_{\rm out}$ and M${\star}$, implying that strong outflows also seem to be more prevalent in low-mass galaxies and which we interpret as evidence for gravitational suppression of feedback. The comparison between N/O and $\eta$ is shown in the bottom panel of Fig.~\ref{fig:outflows}, color-coded by M$_{\star}$. We find a strong correlation between N/O and $\eta$ (Kendall's $\tau = 0.317$, $p$-value $> 10^{-3}$), suggesting that feedback from stellar winds becomes relatively weak as N/O increases. Our fit to the data is as follows: 

\begin{equation}
\mathrm{log(N/O)} = (-0.266\pm0.045)\times \eta - (1.198\pm0.038).
 \end{equation}

This correlation could indicate that, at high stellar mass, metals are more likely retained and mixed with the surrounding ISM rather than expelled or recycled \citep[e.g.,][]{roy21}. Even as the feedback efficiency decreases, the velocity of the wind/outflowing gas does increase, which may further facilitate ready mixing.
Overall, low-mass systems are more efficient in launching outflows, likely due to a combination of compactness \citep[by connecting the mass-radius relation and mass-metallicity relation, e.g.,][]{laser01}, which increases the impact of feedback, and of reduced mass, which reduces gravitational suppression of feedback \citep[e.g.,]{Muratov15, McQuinn19, Llerena23}. However, at fixed mass and radius, feedback tends to be more efficient at high metallicity as line driving is a dominant mechanism in stellar winds \citep[e.g.,][]{leitherer14, byrne22}. See Sec.~\ref{sec:discussion} for more discussion.  

\begin{figure}
\begin{center}
    \includegraphics[width=0.43\textwidth, trim=30 0 30 0,  clip=yes]{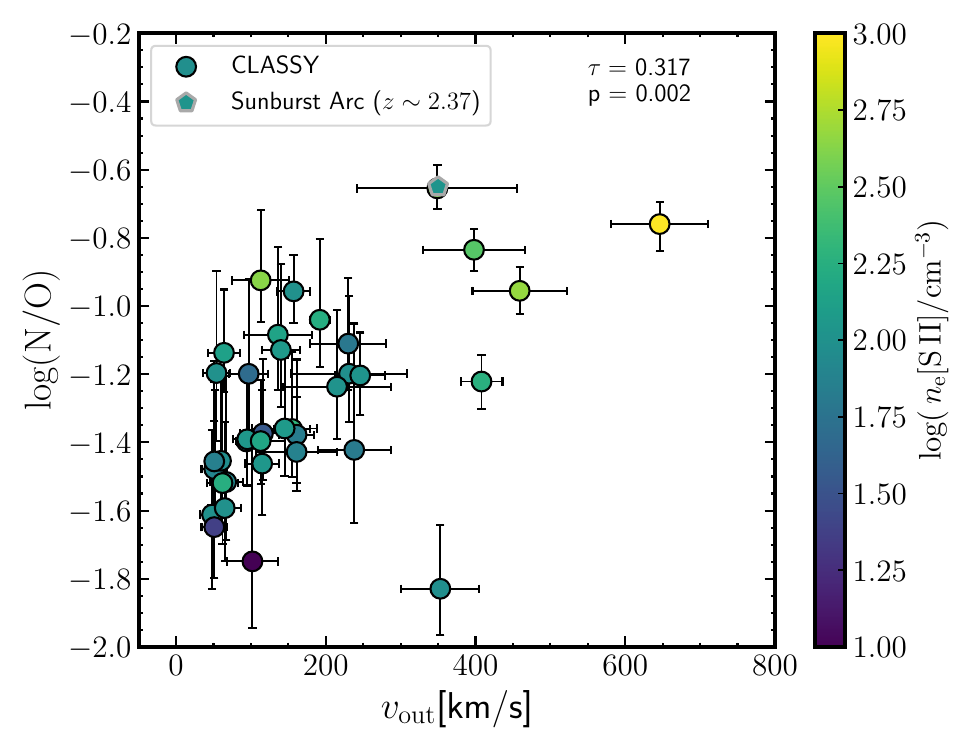}
    \includegraphics[width=0.43\textwidth, trim=30 0 30 0,  clip=yes]{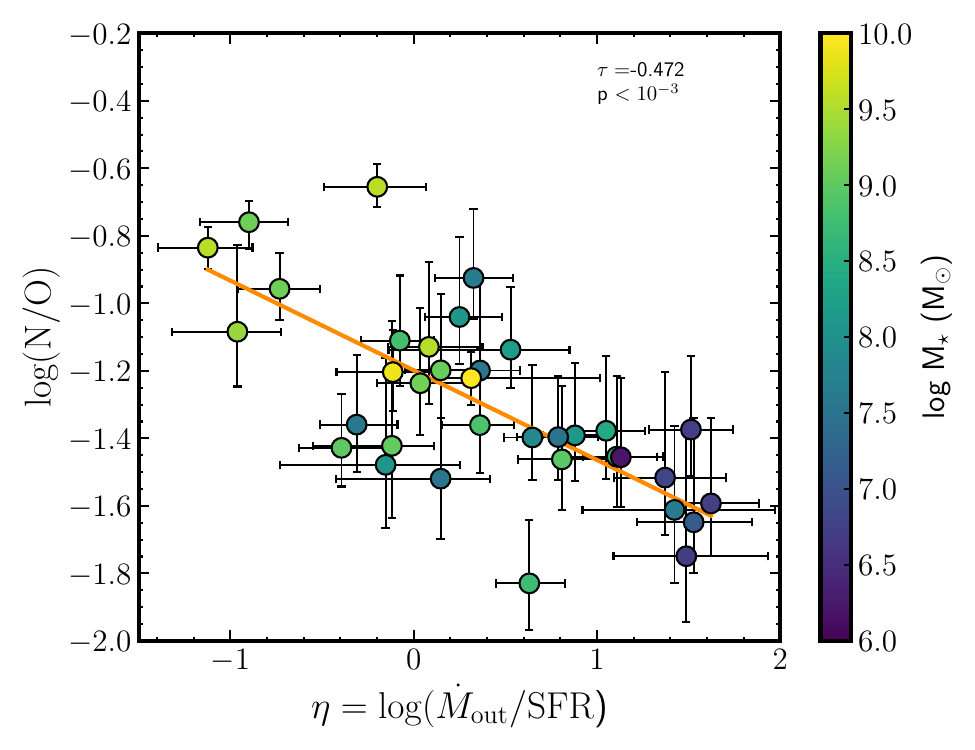}
        \caption{{\it Top:} N/O as a function of outflow velocity ($v_{\rm out}$) \citep{xu22}, color-coded by $n_{\rm e}$ for CLASSY. The Sunburst Arc galaxy is included for comparison with galaxies at $z \sim 0$. The outflow velocity and $n_{\rm e}$ values for Sunburst Arc are from \citet{maianali22} derived from both high ionization absorption and emission lines), while N/O is from \citet{welch25}. {\it Bottom:} N/O as a function of mass-loading factor ($\eta$), color-coded by stellar mass. The Kendall's $\tau$ coefficient and $p$-value are labeled in each plot. The solid line shows the best fit to the data. At high N/O, stellar feedback is stronger but less efficiently produced by young stellar populations.}
\label{fig:outflows}
\end{center}
\end{figure}

\section{Discussion}\label{sec:discussion}
We have analyzed the gas properties and N/O and N/H abundances for the CLASSY survey high$-z$ analogues based on \Te\ abundance determinations, and compared these results with high-$z$ galaxies. Overall, we find that the CLASSY galaxies follow the trend of the N/O-O/H relation observed in other galaxies and \ion{H}{ii} regions at $z\sim0$ with a significant scatter (see Fig.~\ref{fig:NO_highz}). In addition, we identify seven galaxies hosting WR stars, whose log(N/O) values ($< -1.00$) are consistent with those of other galaxies at similar metallicities with no N/O enhancement, except for J0127-0619 \citep[Mrk~996][]{james09}. 
The large scatter in the N/O-O/H relation is well known and can be understood in terms of the different timescales over which nitrogen and oxygen are released into the ISM \citep[e.g.,][]{perez-montero09, berg19, berg19a} and different star formation histories \citep[e.g.,][]{guo16}. Other galaxy properties may also play a crucial role in explaining part of this scatter. For example, \citet{vincenzo16}, using chemical evolution models, show that the SFE significantly influences oxygen production by massive stars. If the SFE is low,  small amounts of oxygen will be ejected to the ISM, implying an increase of the N/O abundance due to the production of N from different generations of stars \citep[][]{berg19a, Schaefer20}. 
However, as shown in Fig.~\ref{fig:NO_highz}, variations in SFE ($\nu = 0.5$-5 Gyr$^{-1}$) do not fully explain the N/O scatter observed in the CLASSY galaxies with the selected chemical evolution models of  \citep[e.g.,][]{vincenzo16}. We have analyzed different galaxy properties that can have an important impact on the scatter around the N/O-O/H relation. In Appendices~\ref{A} and~\ref{B}, we provide additional analysis and discussion on the comparison between N/O and extinction, stellar ages, as well as comments on specific CLASSY galaxies related to gas kinematics and electron density. In the following sections, we analyze our main results from the previous sections.

\subsection{The impact of electron density on N/O}
In principle, the N$^{+}$/O$^{+}$ ratio is not sensitive to variations in \Te\ and \Ne\ due to similar ionization potentials of these ions and their dependence on the critical densities in the low \Ne\ regime \citep{juandedios16}. The variations of density ionized regions has been investigated by \citet[][]{mendez-delgado23b} in a large sample of galaxies and \ion{H}{ii} regions. These authors found that, in the presence of density fluctuations or multiple clumps of dense gas, the O$^{+}$ abundance is underestimated because \Ne[\ion{S}{ii}] or \Ne[\ion{O}{ii}] is insensitive to such high density clumps due to the desity regime that those diagnostics are valid \citep{mendez-delgado+23a}.   

Similarly, our results indicate that high density gas traced by [\ion{S}{ii}] is associated with elevated N/O values (see Fig.\ref{fig:density-structure}). One possible explanation for this positive correlation between N/O and $n_{\rm e}$ is an underestimate of the ionic O$^{+}$ abundance derived from the [\ion{O}{ii}]$\lambda\lambda$3726,29 lines in high density gas \citep[e.g.,][]{stasinka23, Zhang25, mendez-delgado23b}. This underestimation arises from the low critical densities of [\ion{O}{ii}]~$\lambda\lambda$3726,29 ($n_{\rm e, crit} = 10^{2}$-$10^{3}$ cm$^{-3}$). On the other hand, \citet{Zhang25} use photoionization models to analyze the impact of varying \Ne\ =100 cm$^{-3}$ - 20000 cm$^{-3}$ on the UV and optical line ratios commonly used to determine N/O (e.g., [\ion{N}{ii}]/[\ion{O}{ii}]). The study of \citet{Zhang25} showed that UV emission lines (i.e., \ion{N}{iv}] and [\ion{N}{iii}]) are less sensitive to density changes than optical lines, while that high density models predict higher values of [\ion{N}{ii}]/[\ion{O}{ii}] with a difference of 0.2 dex. While our CLASSY sample includes only a few objects at high density, a larger sample of galaxies with densities derived from both [\ion{O}{ii}] and [\ion{S}{ii}] would provide more information into the relationship between N/O and $n_{\rm e}$ at $\log(n_{\rm e}/\mathrm{cm}^{-3}) > 2.5$.

Overall, our results indicate that $n_{\rm e}$ play an important role in shaping N/O, potentially contributing to part of the scatter observed in Fig.~\ref{fig:NO_highz}. Since O$^{+}$ contributes to the total gas metallicity, it is also important to understand the bias introduced at high densities \citep[e.g.,][Martinez et al. 2025]{yanagisawa24b}. The impact of density on metallicity estimates has also been noted in other works, for example \citet{hayes25} shows, for a sample of stacked spectra of $\mathrm{z\sim 4-10}$ galaxies, that the assumption of very high density gas ($10^{6}$ cm$^{-3}$) to infer \TO\ based on the [\ion{O}{iii}]\W5007/~\W1666 ratio can lead to lower \Te\ values that align with those in local galaxies. In those high density environments, [\ion{O}{iii}]\W5007 is collisionally de-excited due to its critical density ($n_{\rm e_{crit}}$ $\sim$ 10$^{5}$ cm$^{-3}$), resulting in a reduction of the inferred \TO, which impacts the metallicity by shifting it to higher O/H values. Therefore, the assumption of high density increases the derived metallicity and decreases N/O; in this case, the elevated N/O values seen in some SFGs at high redshift might at least in part be explained by incorrect density assumptions (Martinez et al. 2025). Note that the N/O derived in \citet{hayes25} is based on UV N lines, which are less sensitive to density variations \citep[e.g.,][]{marqueschaves24, Zhang25}. 

On the other hand, some studies have analyzed the redshift evolution of \Ne [\ion{S}{ii}] \citep[e.g.,][]{sanders16, isobe23a, topping25} finding densities higher than those typical local SFGs. Recently, \citet{topping25} use a sample of galaxies from the AURORA survey \citep{shapley24}, finding a  correlation between \Ne[\ion{S}{ii}] and redshift in galaxies at $z=2-6$ with average values between 300~cm$^{-3}$ and 500~cm$^{-3}$, similar to the values reported for CLASSY galaxies \citep[][]{berg22, mingozzi22, arellanocordova24}.  Our analysis with CLASSY also suggests that high density {gas} are associated with an enhancement of N/O, independent of the metallicity of the gas \citep[e.g.,][]{flury24a, yanagisawa24b}. 

\subsection{The impact of $M_{\star}$ and SFR on N/O}

The relation between N/O with stellar mass has previously been studied using samples of local and high-$z$ galaxies \citep[e.g.,][]{perez-montero09, amorin10,  strom17, andrews13, perezmontero13,hayden-pawson22}. In \citet{perez-montero09}, these authors show that more massive galaxies implied high N/O abundance than the lower counterpart because of the rapid chemical evolution of a more massive system. \citet{hayden-pawson22} show that $z\sim2$ galaxies are systematically 0.35 dex shifted to lower N/O in comparison with local galaxies. In contrast, other studies predicted a shift to elevated N/O at higher redshifts \citep[e.g.,][]{masters16}. However, note such studies are based on the use of strong-line methods in both local and higher redshift observations.

In Fig.~\ref{fig:NO_scatter_properties}, we show that the variations in N/O are across stellar masses for CLASSY with large scatter. To analyze the redshift evolution of the N/O and stellar mass plane, we have added the sample of high$-z$ galaxies that we have gathered from the literature. We found that the $z=2-6$ sample, whose N/O derived by optical lines is in good agreement with CLASSY ($z\sim0$) showing no apparent evolution for a fixed stellar mass up to $z\sim6$. However, $z>6$ galaxies (UV N) depart significantly from the bulk of the sample for a fixed stellar mass. For N/H, we get similar results when we include $z=2-6$ galaxies.
In general, when we compare the N/H-M$_{\star}$ plane, our results do not indicate a redshift evolution.

To understate what is causing the dispersion on the N/O -M$_\star$ (see  the top panel of Fig.~\ref{fig:NO_scatter_properties}), we compare such a relation in color-codded with \Ne[\ion{S}{ii}] in Fig.~\ref{fig:SFR_mass_density}. Overall, we find that the high density galaxies are clearly causing the scatter for a fixed stellar mass. 
In addition, we have incorporated the same chemical evolution models as in Fig.~\ref{fig:NO_highz}, but as a function of stellar mass in Fig.~\ref{fig:SFR_mass_density}, illustrating different SFE. For comparison, we also show the models used by \citet{hayden-pawson22} in their study of N/O at $z\sim2$. The chemical evolution models presented in \citet{hayden-pawson22} assume a \citet{salpeter55} IMF, a higher infall mass of log(M${\rm inf}$/M$\odot$) = 12, and a SFE of $\nu = 5$ Gyr$^{-1}$ and $\tau_{\rm inf} = 7$ Gyr \citep[see also][]{vincenzo16}.
While most of the CLASSY galaxies lie within the parameter space for a fixed stellar mass, high density galaxies deviate from the region traced by the models. Notably, galaxies at $z = 2-6$ also align with the set of chemical evolution models presented here, in contrast to the selected models from \citet{hayden-pawson22} that include high time infall ($\tau_{\rm inf} = 7$ Gyr) and log(M$_{\rm infall}$/M$\odot$)$=12$, which were used to cover their sample of $z \sim 2$ galaxies. Overall, our results confirm that N/O increases with stellar mass in more evolved galaxies (high M$_\star$) \citep[e.g.,][]{perezmontero13}, but also shows a significant scatter associated with high density gas at fixed stellar mass.

Another essential property in the context of galaxy evolution is star formation, we show in Fig.~\ref{fig:NO_scatter_properties} that N/O increases as SFR increases for values of log(SFR/M$_{\odot}$yr$^{-1}$) $\ga1$. Although high values of N/O cover a wide range of stellar mass, the CLASSY galaxies are more closely correlated with the SFR. Such a result is also in agreement with other galaxies at $z\sim0$ (grey circles). Interestingly, when we include the sample of galaxies at high $z=2-6$ (pentagons), we find that these high$-z$ galaxies also follow the relation for CLASSY ($z\sim0$) with no apparent redshift evolution.  We have also inspected the location of $z>6$ galaxies (squares). Overall, we find that the comparison between N/O and SFR led to a tight relation with less scatter as the SFR increases. In the bottom panel of Fig.~\ref{fig:SFR_mass_density}, we show the N/O-SFR plane in color-codded with \Ne, which shows most of the CLASSY galaxies with (log\Ne[\ion{S}{ii}]/cm$^{-3}$)$>2.5$) are mainly associated with high N/O and high SFGs. In a similar way, high-$z$ galaxies tend to follow a similar dependency with \Ne\ than local galaxies. Our results in Figs.~\ref{fig:NO_scatter_properties} and~\ref{fig:SFR_mass_density} indicate that \Ne\ and high SFR might contribute to the scatter of the N/O-O/H and the stellar mass, and the enhancement of N/O for a fixed metallicity.

\begin{figure}
\begin{center}
    \includegraphics[width=0.48\textwidth,  clip=yes]{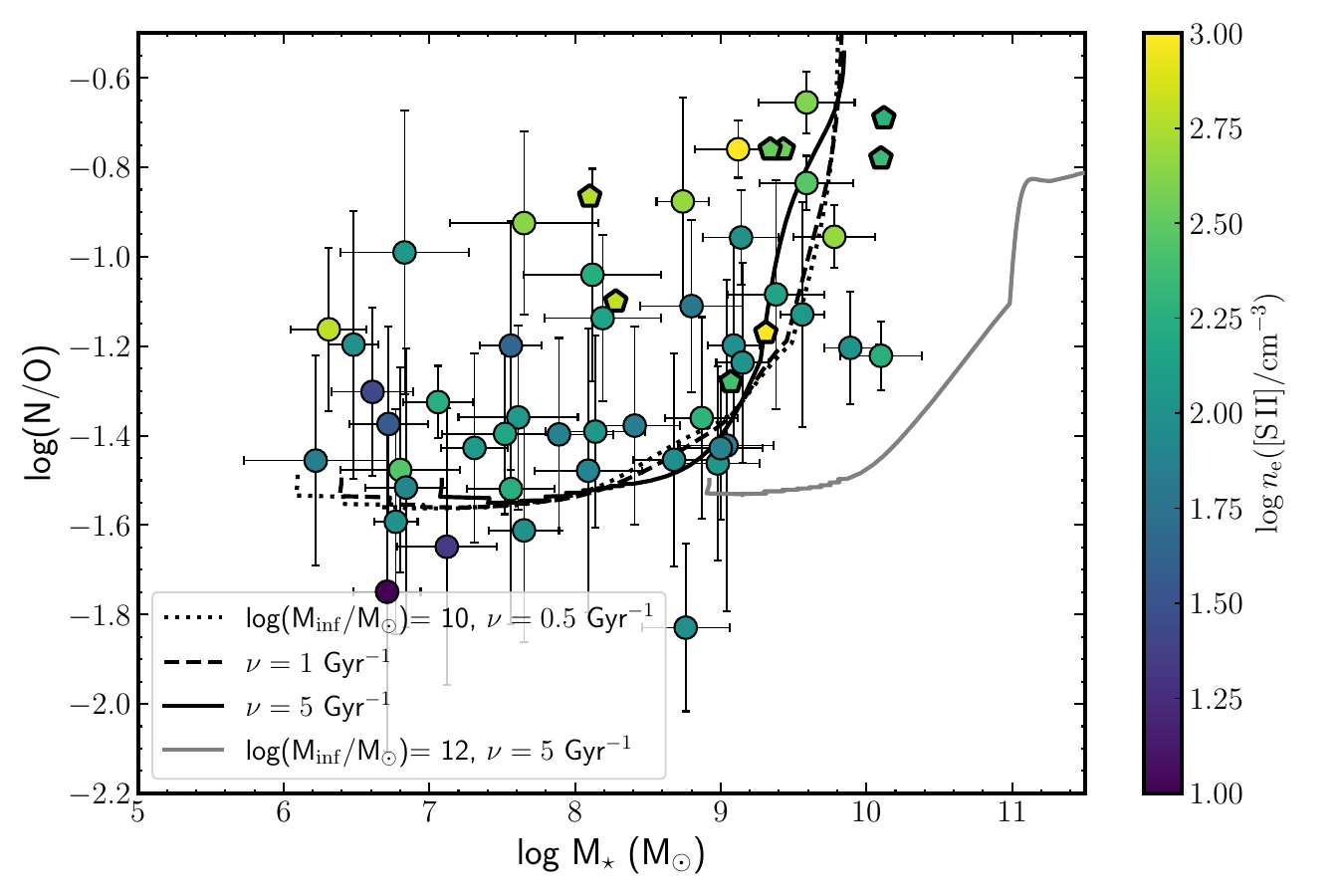}
    \includegraphics[width=0.48\textwidth, trim=0 0 0 0,  clip=yes]{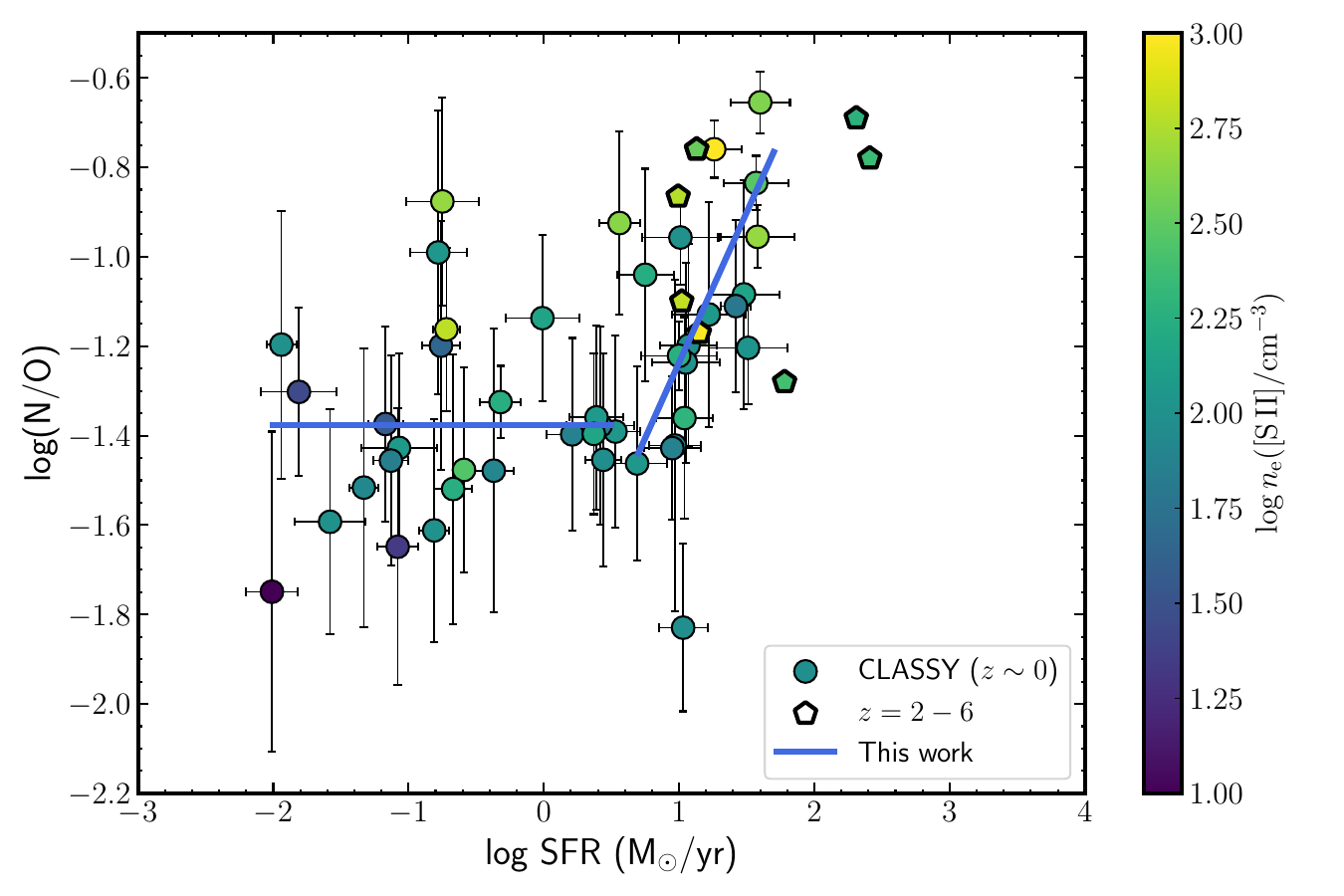}
        \caption{Comparison of N/O as a function of stellar mass (top) and SFR (bottom), color-coded by \Ne[\ion{S}{ii}]. Part of the dispersion in these relations is due to galaxies exhibiting high density and elevated N/O ratios. Most galaxies at $z = 2-6$ are also located in a similar parameter space in N/O and \Ne\ as those in the CLASSY sample (see also Fig.~\ref{fig:density-structure}). The different lines represent the chemical evolution models as in Fig.~\ref{fig:NO_highz}, varying only the SFE ($\nu$). For comparison, we also include the chemical evolution models from \citet{hayden-pawson22} (gray line), which cover the parameter space of their galaxy sample at $z \sim 2$. The offset in stellar mass is due to higher value in M$_{\rm inf}$ assuming by \citet{hayden-pawson22}.  Note that most of the galaxies at $z = 2-6$ (pentagons) also follow the models presented in this study.} 
\label{fig:SFR_mass_density}
\end{center}
\end{figure}

Since SFGs at high N/O tend to be very compact \citep[e.g.,][]{schaerer24, Adamo24, Ji25}, we also investigated the connection between N/O and both SFR and stellar mass surface densities (see Fig.\ref{fig:SFR_density}). Overall, our results show a tentative correlation between N/O and $\Sigma_{\rm SFR}$, while no clear correlation is found with $\Sigma_{M_\star}$ (see Sec.\ref{sec:Galaxy_properties}). Since electron density appears closely linked to high N/O values, we selected the CLASSY galaxies with $n_{\rm e}$([\ion{S}{ii}]) > 300 cm$^{-3}$ and compared them with galaxies at $z = 2-6$. In Fig.~\ref{fig:SFR_mass_Density_high_z}, we compare N/O with $\Sigma_{\rm SFR}$ (left) and $\Sigma_{M_\star}$ (right) for the selected high density CLASSY galaxies, color-coded by metallicity. We find that, as expected, the CLASSY galaxies cover a lower range of $\Sigma_{\rm SFR}$ and $\Sigma_{M_\star}$, with metallicities spanning 12 + log(O/H) = 8.0-8.76, but they show N/O values similar to those of galaxies at $z = 2-6$.

Unfortunately, our comparison sample at $z = 2-6$ includes only three galaxies with N/O inferred from optical lines, as their effective radii are often not reported in the original studies. In addition, we also include $z > 6$ galaxies where $n_{\rm e}$ is determined from UV lines (e.g., \ion{C}{iii}] or \ion{Si}{iii}]). Note that  $z > 6$ galaxies exhibit SFRs similar to those at lower redshift with comparable chemical abundance patterns (e.g., Ne/O) \citep[e.g.,][]{arellanocordova22b, arellanocordova25, stanton2025}. 
As we see in Fig.\ref{fig:SFR_mass_Density_high_z}, $z > 6$ galaxies show elevated $\Sigma_{\rm SFR}$ and $\Sigma_{M_\star}$, as seen in other recent studies \citep[e.g.,][]{Ji25}. Fig.\ref{fig:SFR_mass_Density_high_z} also shows that galaxies with similar N/O cover different ranges of $\Sigma_{\rm SFR}$ and $\Sigma_{M_\star}$. For one galaxy at $z \sim 6$, RXCJ2248-ID, it is possible to determine the N/O ratio using only optical lines. \citet{arellanocordova25} inferred the optical N/O ratio of RXCJ2248-ID from the physical conditions and corrected fluxes of [\ion{N}{ii}] and [\ion{O}{ii}] reported by \citet{topping24}, reporting a high log(N/O) = $-0.60 \pm 0.15$ compared to the local abundance, which indicates that both UV and optical N lines show a similar degree of enhancement (see also Sec.\ref{sec:density}). This value is illustrated in Fig.~\ref{fig:SFR_mass_Density_high_z} for comparison (rhomboid).
Overall, the CLASSY galaxies with high N/O and high \Ne\ are those with relatively higher metallicity compared to higher-$z$ galaxies. This is consistent with the expectation that, in more evolved galaxies, the contribution of AGB stars enriches the ISM with secondary nitrogen production at 12 + log(O/H) $\gtrsim 8.0$ \citep[e.g.,][]{vincenzo16}. Additionally, our results suggest that part of the dispersion in the N/O-O/H relation (see Fig.\ref{fig:NO_highz}) might be driven by other properties, such as high log($\Sigma_{\rm SFR}$/M${\odot}$ yr$^{-1}$ kpc$^{-2}$) $> 1$ in objects that also have high \Ne([\ion{S}{ii}]) $\gtrsim 300$ cm$^{-3}$ (see Figs.~\ref{fig:SFR_mass_density} and~\ref{fig:SFR_mass_Density_high_z}). Such a trend might also extend to galaxies at $z=2-6$.

\subsection{The impact of stellar feedback on N/O}

The baryon cycle is the key link between stellar processes that produce heavy elements in galaxies. In this context, an important component of the baryon cycle is galactic outflows. Our analysis in Fig.~\ref{fig:outflows} indicates that strong outflows correlate with N/O enhancement at $z \sim 0$. 

Recent studies of the Sunburst Arc, a galaxy at $z = 2.37$ (12 + log(O/H) = 7.97), have revealed interesting physical properties. For example, the analysis of one of its massive clusters shows N/O enhancement likely related to the presence of WR stars \citep{riverathorsen24, welch25, pascale23}. Our analysis at $z \sim 0$ shows that WR enrichment has a minimal impact on N/O abundances; however, our small sample covers also higher metallicities than the Sunburst Arc. Additionally, \citet{maianali22} report a strong outflow velocity derived from UV absorption lines ($v_{\rm out} = 350\rm~km~s^{-1}$) and $n_{\rm e}$([\ion{O}{ii}]) $\sim 300$ cm$^{-3}$, which aligns with the results for galaxies at $z \sim 0$ (see Fig.~\ref{fig:outflows}). This suggests that feedback from massive stars also plays an important role in nitrogen enhancement at $z \sim 2$. Another interesting result from this analysis is the anticorrelation between $\eta$ and outflow velocities (bottom panel in Figure 7). 
The expectation from chemical evolution models is that oxygen is preferentially evacuated in outflows relative to other elements \citep[e.g.,][]{vincenzo16, Bhattacharya25}, which suggests a net positive correlation between $\eta$ and N/O. This contrasts with what we find in CLASSY. Our results could indicate that nitrogen produced by massive stars is in fact incorporated into outflows \citep[e.g.,][]{Rizzuti25}. This effect might be due to the fact that these same massive stars and their surrounding environments are associated with the subsequent CCSNe which drive outflows. However, because the mass threshold for SNe onset may be much lower at lower metallicities \citep{jecmen23}, the production and ejection of O as well as the mass loading should be lower in low-metallicity galaxies if SNe are driving the observed outflows. As such, stellar winds may be the key driving mechanism of the observed kinematics. Thus, it seems more likely that the inverse relation between N/O and $\eta$ is related to N-rich, oxygen poor winds launched by massive stars. While gravitational suppression reduces the outflow launching efficiency, the wind speeds are higher due to line driving (see Sec.~\ref{sec:feedback}), which could explain the observed correlation. N-rich winds explain not only the observed correlation with $\eta$ but also the trend with density. For instance, \citet{flury24} suggested that 100-200 $\rm km~s^{-1}$ winds or outflows could trigger shocks that enhance gas density. They find that their shock scenario is consistent with galactic winds containing N/O excesses of 0.5 dex.

\begin{figure*}
\begin{center}
    \includegraphics[width=0.95\textwidth, trim=0 0 0 0,  clip=yes]{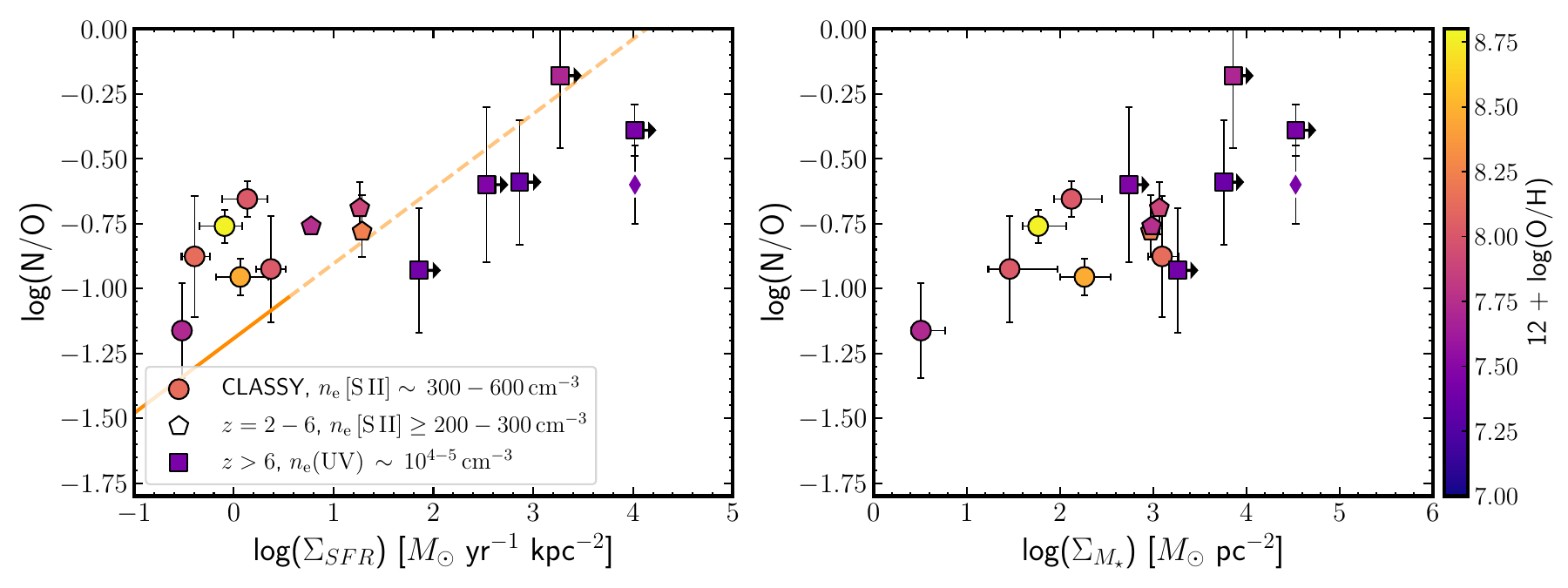}
        \caption{The $\Sigma_{SFR}$ and $\Sigma_{M_{\star}}$ as a function of N/O for high density CLASSY galaxies and high-$z$ galaxies. The right panel shows the ranges in \Ne\ for each sample. The sample of high-redshift galaxies is also included and distinguished according to whether N/O is derived using UV \citep[squares,][]{marqueschaves24, topping24, topping24b, schaerer24, curti24b} or optical \citep[pentagons,][]{sanders23, Zhang25} N lines as in Fig~\ref{fig:NO_highz}. For RXCJ2248-ID, we have included the N/O ratio derived using optical (diamond symbol) by \citet{arellanocordova25} for comparison with the UV N/O derived in \citet{topping24}. The CLASSY galaxies with high density values tend to have similar N/O enhancements as galaxies at $z = 2-6$, but exhibit lower star formation rate and stellar mass surface densities. The combined solid and dashed lines represent an extension of the relation between N/O and $\Sigma_{SFR}$ derive for CLASSY at $z\sim0$. Such a relation suggest that N/O increases as the compactness of the galaxy increases.}
\label{fig:SFR_mass_Density_high_z}
\end{center}
\end{figure*}

\subsection{Chemical evolution models for N and O}

By comparing observed abundance patterns with predictions from chemical evolution models, we can place direct constraints on the sources and pathways of enrichment \citep[e.g.,][]{vincenzo16, berg19a, kobayashi20}. Here, we investigate the scatter in the N/O-O/H relation by developing new models in the low-metallicity regime (12 + log(O/H) $\la$ 8.0) that consider a simple chemical enrichment scenario. In this regime of low metallicity, the CLASSY galaxies, as well as other dwarf galaxies (gray circles) from the literature, show high log(N/O) $> -1.25$ with respect to the bulk of the observations (see Fig.~\ref{fig:NO_highz}). From our findings in Sec.~\ref{sec:feedback}, we show that low-mass galaxies (most of them located in the plateau of the N/O-O/H relation) with high N/O might undergo relatively strong feedback from massive stars.

Our models assume a stellar cluster with a total mass of $10^4$ M$_{\odot}$, adopting the Kroupa IMF \citep{kroupa93}. We include rapidly rotating massive stars with initial masses in the range 13–130 M$_\odot$ and adopt the stellar yields from \citet{limongiandchieffi18} with rotation velocity $v_{\text{rot}}=300\,\text{km/s}$. In these models, we do not include contributions from AGB stars. Every time a massive star is formed in the IMF sampling, the model updates the cumulative mass of O and N that is ejected in the ISM until a stellar cluster mass of $10^4$ M$_{\odot}$ is formed. A key parameter in these models related to feedback is $f_{\rm retained}$, the fraction of explosive yields retained in ISM from massive stars with $M < 25$ M$_\odot$. We assume that all stellar wind material is fully retained, independent of stellar mass or metallicity, whereas how much is retained from the explosion of stars with $M < 25$ M$_\odot$ is regulated by $f_{\rm retained}$. The selected threshold of $M < 25$ M$_\odot$ is based both on observations of the mass distribution of CCSNe progenitors \citep[e.g.,][]{smartt09, oconnor11} and on the explodability of massive stars and/or black hole formation from stellar evolution models \citep[e.g.,][]{Sukhbold16}.

In Fig.~\ref{fig:modelos_evolution}, we present the results of our chemical enrichment models in comparison with low metallicity CLASSY galaxies. For these low-metallicity galaxies, we find a mean and standard deviation of log(N/O) = $-1.31 \pm 0.23$ (dashed line), which is slightly elevated compared to our model predictions (colored diamonds) and the values observed in other dwarf galaxies (gray circles).
Our models indicate that most CLASSY galaxies are consistent with scenarios where $f_{\rm retained} > 0.6$, suggesting that a significant fraction of the supernova ejecta is retained.

However, the CLASSY galaxies exhibiting higher than average N/O ratios may be explained by a more dominant contribution from stellar winds to N enrichment. In Fig.~\ref{fig:outflows}, we report relatively strong outflows in low-mass, low-metallicity galaxies compared to galaxies with high stellar masses and metallicities \citep[see also][]{xu22}. With this correlation in mind, we interpret our model predictions for galaxies with high N/O and low $f_{\rm retained} \sim 0.2$ as indicative of enrichment dominated by stellar winds, with only a small fraction of the explosive ejecta being retained. A follow-up analysis of the N/O abundance patterns in low-mass galaxies is essential to better constrain the role of stellar feedback, including both stellar winds and supernovae, in regulating nitrogen enrichment.

\begin{figure}
\begin{center}
    \includegraphics[width=0.5\textwidth, trim=0 0 0 0,  clip=yes]{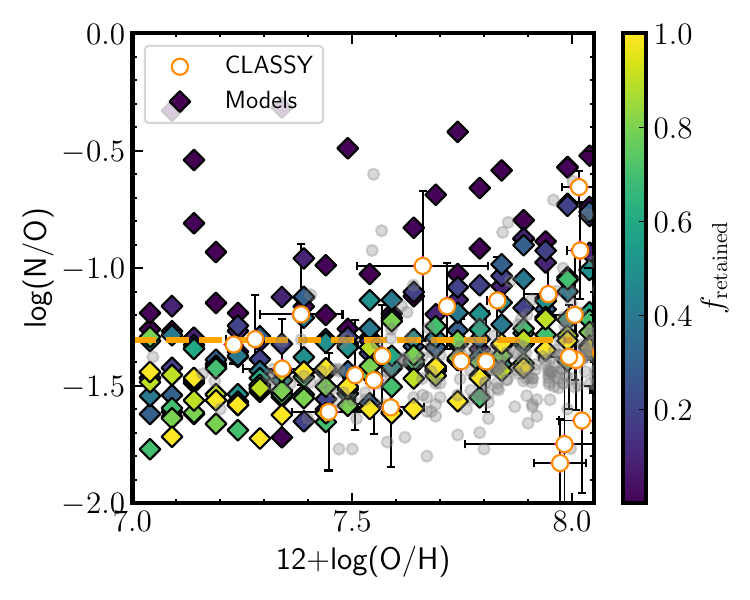}
        \caption{ Predicted chemical abundance patternfor a stellar cluster with a total mass of 10$^4$ M$_\odot$ in the N/O–O/H relation, color-coded by the retained \textbf{explosive} yield fraction ($f_{\rm retained}$). CLASSY galaxies and other local star-forming galaxies (SFGs) are included for comparison. The dashed line represents the mean and standard deviation of log(N/O) = $-1.31 \pm 0.23$ for CLASSY galaxies with 12 + log(O/H) $<$ 8.0.}
\label{fig:modelos_evolution}
\end{center}
\end{figure}

\section{Summary and Conclusions} 
\label{sec:conclusion}
We investigate the chemical evolution of nitrogen along with oxygen using a sample of 45 local SFGs from the CLASSY survey. This sample includes a broad range of galaxy properties with robust chemical abundance determinations of N and O via the direct-\Te\ method. We compare the N/O abundance in CLASSY galaxies with their electron density and various properties such as stellar age, stellar mass, SFR, and kinematics (see Sec.~\ref{sec: sample}).  Additionally, we present an analysis of the SFR and stellar mass surface density and their relationship with the N/O ratio across cosmic time. For comparison with our CLASSY findings, we have also included a sample of ($z=2-10$) galaxies where N/O ratios were determined using UV or optical N lines. In this study, we seek to understand the chemical enrichment pathways of N/O at different metallicities, using a well-defined sample that mimics the properties of high-$z$ galaxies. Moreover, we investigate the main physical drivers of the scatter in the N/O--O/H ratio at $z\sim0$, to guide the interpretation of high-redshift galaxies observed with JWST or other current and future facilities. Our main conclusions from this study are as follows:

\begin{enumerate}
    
     \item[--] We present the N/O-O/H relation for the CLASSY survey, covering a wide range of metallicity values (12+log(O/H) = 7-9). We find that the N/O abundance ratios align with the expected trend observed in other dwarf galaxies and \hii\ regions at both low and high metallicities, though with some scatter. Particularly, CLASSY show more a constant behavior for a fixed metallicity. 

    \item [--] We found that$z=2-6$ galaxies exhibit the same correlation and dispersion in N/O-O/H as CLASSY galaxies and other systems at $z=0$. This similarity suggests that chemical enrichment and evolution pathways do not change with redshift. I.e., comparison of high redshift with CLASSY galaxies indicates not only shared chemical enrichment patterns but also a complete lack of redshift evolution in the production of N relative to O.
     
    \item[--] We report a significant correlation between \Ne([\ion{S}{ii}]) and the N/O ratio (see Fig.\ref{fig:density-structure}). SFGs with log($n_{\rm e}$/cm$^{-3}$) $> 2.5$ tend to have elevated log(N/O) $\sim -1.0$ (see Fig.\ref{fig:density-structure}). This result is consistent with findings from galaxies at $z = 2-6$, where elevated N/O persists in high-density environments \citep[e.g.,][]{reddy23a, stanton2025}. These results suggest that density structure may significantly contribute to the scatter in the N/O-O/H relation. Therefore, follow-up analyses using galaxy samples with robust density diagnostics are needed to assess the impact of the density structure on the N/O ratio \citep[e.g.,][Martinez et al. 2025]{berg21a, mendez-delgado23b, hayes25, Zhang25}. 
    
   \item[--] Our comparison between N/O and the global properties of CLASSY galaxies shows that SFGs with log(SFR/M$\odot$ yr$^{-1}$) $> 1$ have elevated log(N/O) $> -1.2$ compared to galaxies with lower SFR. Additionally, galaxies with high SFR also exhibit high \Ne([\ion{S}{ii}]) (see Fig.~\ref{fig:SFR_mass_density}). Interestingly, we find large scatter in N/O as a function of stellar mass, although the correlation is weak. To investigate the origin of this dispersion, our analysis suggests that the density structure contributes, at least in part, to the scatter at fixed M$_\star$. Notably, $z = 2-6$ galaxies occupy a similar parameter space to that implied by CLASSY galaxies at $z \sim 0$ with no apparent redshift evolution.

    \item[--] We investigate whether the chemical enrichment of N/O correlates with stellar mass and SFR surface density. We find a moderate correlation between compactness and high N/O, but no clear correlation with stellar mass surface density at $z \sim 0$. Since $\Sigma_{\rm SFR}$ tends to correlate with $n_{\rm e}$, we compare CLASSY galaxies with high $n_{\rm e}$([\ion{S}{ii}]) to high-$z$ galaxies. We find that galaxies at $z = 2-6$ show similar N/O values but with more compact star formation, which may be consistent with our finding of no chemical evolution in the N/O-O/H relation. In contrast, galaxies at $z > 6$ with more extreme compact star formation are metal-poor, suggesting that, in addition to compactness, other physical mechanisms may play an important role in driving the extreme N/O ratios.

   \item[--] From the analysis of N/O and gas kinematics from UV absorption lines, we find that high velocity outflows ( >100 $\rm km~s^{-1}$) are associated with elevated N/O and high $n_{\rm e}$. We also find that the mass-loading factor, $\eta$, strongly anticorrelates with N/O, indicating that galaxies with higher N/O (and high stellar mass) are associated with weaker feedback by massive stars. We do not find any clear correlation among N/O and $\Sigma_{\rm SFR}$, suggesting that compactness is a property that allows feedback to have a bigger impact for chemical enrichment of N/O at $z\sim0$. However, at $z \sim 3$, galaxies tend to be more compact, and previous studies have shown that compact star formation correlates with higher values of $\eta$ (derived using emission lines) \citep[e.g.,][]{Llerena23}, in contrast to what we observe in the CLASSY sample. These differences may suggests changes in the relationship between outflows and chemical enrichment. However, \citet{Xu25} shows the mass-loading factor derived from UV absorption lines is systematically larger than the $\eta$ derived from emission lines.
   \end{enumerate}

The results presented here show that high \Ne[\ion{S}{ii}], high SFR, high outflow velocities, and reduced feedback relative to the stellar populations together play a crucial role in shaping the observed N/O values. Interestingly, galaxies at $z = 2$–6 follow the local trend seen in CLASSY, showing no apparent redshift evolution. Larger samples at high $z$ with diverse density diagnostics are essential to improve our understanding of the interplay of all these physical properties in setting N/O levels within galaxies at $z \sim 0$. The scatter in the N/O–O/H relation may reflect a combination of density, SF, and outflows tracing different gas conditions. These findings highlight the need for joint constraints on feedback, chemical abundances, and ISM conditions to understand the chemical evolution of nitrogen in galaxies across cosmic time. A follow-up analysis requires a sample of galaxies with high-resolution UV and optical spectra to connect gas outflow properties across different kinematic components with their chemical enrichment. 


\section*{Acknowledgments}
The CLASSY team thanks to our referee Igor Zinchenko for thoughtful feedback that improved this paper.
KZA-C and DAB are grateful for the support for this program, HST-GO-15840, that was provided by 
NASA through a grant from the Space Telescope Science Institute, which is operated by the Associations of Universities for Research in Astronomy, 
Incorporated, under NASA contract NAS5-26555. 
The CLASSY collaboration extends special gratitude to the Lorentz Center for useful discussions 
during the "Characterizing Galaxies with Spectroscopy with a view for JWST" 2017 workshop that led 
to the formation of the CLASSY collaboration and survey.
The CLASSY collaboration thanks the COS team for all their assistance and advice in the 
reduction of the COS data. 

KZA-C and FC acknowledge support from a UKRI Frontier Research Guarantee Grant (PI Cullen; grant reference: EP/X021025/1). 
MGS acknowledges support from NSF REU grant AST-1757983 (PI: Jogee) funded by the NSF and Department of Defense, and for support from REU mentors KZA-C and DAB.
RA acknowledges support of Grant PID2023-147386NB-I00 funded by MICIU/AEI/10.13039/501100011033 and by ERDF/EU, the Severo Ochoa award to the IAA CEX2021-001131-S, and from the grant PID2022-136598NBC32 “Estallidos8”.

MM, BLJ, SH, NK are thankful for support from the European Space Agency (ESA).

CK acknowledges funding from the UK Science and Technology Facility Council through grant ST/Y001443/1.

This work also uses observations obtained with the Large 11 Binocular Telescope (LBT). The LBT is an international collaboration among institutions in the United States, Italy and Germany. LBT Corporation partners are: The University of Arizona on behalf of the Arizona Board of Regents; Istituto Nazionale di Astrofisica, Italy; LBT Beteiligungsge-sellschaft, Germany, representing the Max-Planck Society,The Leibniz Institute for Astrophysics Potsdam, and Heidelberg University; The Ohio State University, University of
19 Notre Dame, University of Minnesota, and University of Virginia.

Funding for SDSS-III has been provided by the Alfred P. Sloan Foundation, the Participating Institutions, the National Science Foundation, and the U.S. Department of Energy Office of Science. 

The SDSS-III web site is http://www.sdss3.org/.
SDSS-III is managed by the Astrophysical Research Consortium for the Participating Institutions of the SDSS-III Collaboration including the University of Arizona, the Brazilian Participation Group, Brookhaven National Laboratory, Carnegie Mellon University, University of Florida, the French Participation Group, the German Participation Group, Harvard University, the Instituto de Astrofisica de Canarias, the Michigan State/Notre Dame/JINA Participation Group, Johns Hopkins University, Lawrence Berkeley National Laboratory, Max Planck Institute for Astrophysics, Max Planck Institute for Extraterrestrial Physics, New Mexico State University, New York University, Ohio State University, Pennsylvania State University, University of Portsmouth, Princeton University, the Spanish Participation Group, University of Tokyo, University of Utah, Vanderbilt University, University of Virginia, University of Washington, and Yale University.

This work also uses the services of the ESO Science Archive Facility,
observations collected at the European Southern Observatory under 
ESO programmes 096.B-0690, 0103.B-0531, 0103.D-0705, and 0104.D-0503.


Software: \texttt{jupyter} \citep{kluyver16}, \texttt{astropy} \citep{astropy:2018, astropy:2022}, {\tt PyNeb} \citep{luridiana15}, {\tt matplotlib} \citep{matplotlib},
{\tt numpy} \citep{numpy}, {\tt scipy} \citep{scipy}, {\tt towrecan} \citep{Flury2025}.

For the purpose of open access, the author has applied a Cre- ative Commons Attribution (CC BY) licence to any Author Ac- cepted Manuscript version arising from this submission.

\section*{Data Availability}
All data will be shared by the corresponding author upon reasonable request.



\bibliographystyle{mnras}
\bibliography{mybib} 




\appendix
\section{Gas and stellar properties}
\label{A}
\subsection{Extinction}
We have also assessed the impact of gas extinctions and galaxy ages on the N/O-O/H dispersion in Fig.~\ref{fig:NO_highz}.
The wavelength separation between the [\ion{N}{ii}] and [\ion{O}{ii}] lines makes the [\ion{N}{ii}]/[\ion{O}{ii}] ratio more sensitive to extinction corrections.
We have compared the $E(B-V)$ values from CLASSY reported in \citet{berg22} to analyze whether the N/O-O/H relation depends on extinction (see also Sec.~\ref{sec: sample}). 
Overall, Fig.~\ref{fig:extiction} shows a tentative positive correlation ($\tau=0.278$ and p-value = 0.08). It implies that the N/O ratio tends to increase with extinction. Interestingly, most CLASSY galaxies with high density, log(\Ne[\ion{S}{ii}]/cm$^{-3}$)$ > 2.5$, are associated with dustier environments than higher redshift galaxies. This result is in agreement with previous analysis in Lyman continuum emitters (LCE) where dense gas prevails in regions of high gas extinction \citep{flury24a}.

We have also incorporated a high-$z$ sample with N/O and \Ne[\ion{S}{ii}] for comparison with the CLASSY results. In Fig.~\ref{fig:extiction}, we show that high-density objects at $z>2$ are located below $E(B-V) < 0.2$ with a shift to high N/O compared to galaxies at $z\sim0$. Recent studies have explored the anomalous discrepancy between the theoretical and observed Balmer ratios based on the assumption of CASE B recombination \citep[e.g.,][]{Yanagisawa24, McClymont24, scarlata+24}. Therefore, any failure in dust correction might lead to incorrect values in the determination of N/O which probably biases our interpretation of Fig.~\ref{fig:extiction}. More samples of galaxies at high-$z$ with high-density and N/O are needed to verify its dependency in dustier environments. 

\begin{figure}
\begin{center}
    \includegraphics[width=0.45\textwidth, trim=40 0 20 0,  clip=yes]{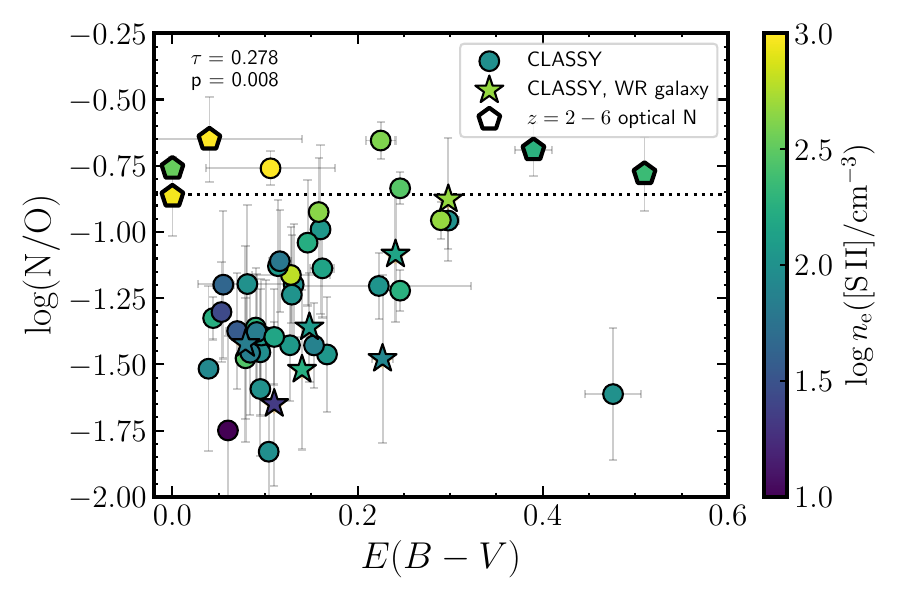}
        \caption{The N/O ratio as a function of the $E(B-V)$ value for CLASSY color-codded against \Ne. There is a tentative positive correlation between N/O and $E(B-V)$, with a dependency with \Ne.  
        High-$z$ galaxies are added for comparison showing a shift to high N/O and \Ne\ \citep{sanders23, arellanocordova25, welch25, stanton2025, scholte25}. The $\tau = 0.278$ and p-value = 0.08 are also labeled. The dotted line indicates the Solar N/O value. The CLASSY galaxies that host WR stars are marked with a star symbol. The dotted line represents the Solar value of N/O.}
\label{fig:extiction}
\end{center}
\end{figure}

\subsection{EW(H$\beta$) and stellar age}
We have revised the EW(H$\beta$) as an empirical indicator for intermediate and old ages of the stellar population. EW(H$\beta$) provides information on the current and previous star formation. Intermediate and old stellar populations do not produce ionizing photons necessary for H$\beta$ emission; however, they can produce and even dominate the underlying continuum. Galaxies with older ages are more evolved, and N/O increases as the new-generation star ejects N into the ISM \citep[e.g.,][]{perez-montero09, vincenzo16, Johnson2023}. Thus, lower EW(H$\beta$) may provide evidence for contributions from older stellar populations to chemical enrichment. However, using EW(H$\beta$) purely as an age diagnostic in this context may be misleading given degeneracies between age and stellar mass. Interpreting the EW is further complicated by the fact that, for sufficiently old stellar populations, no ionizing photons are available to produce H$\beta$ emission, meaning that EW(H$\beta$) traces only populations younger than 10-20 Myr, a timescale too short to correspond to the AGB stars relevant to nitrogen enrichment.

In the top panel of Fig.~\ref{fig:stellar_age_properties}, we show the relationship between EW(H$\beta$) and N/O color-coded with the electron density. For EW(H$\beta$),  we note that the high-N/O is located across all ages with any statistical corelation ($\tau=-0.154$ and $p=0.140$). 
Moreover, Fig.~\ref{fig:stellar_age_properties} also shows that the EW(H$\beta$)-N/O plane depends on density, indicating that density, not stellar population age, drives the observed N/O abundances (see Section \ref{sec:density}). 
In a recent study by \citet{topping25}, these authors reported a significant correlation between EW(H$\beta$) and \Ne[\ion{S}{ii}] using cosmological simulations and observational data from the AURORA Survey \citep{shapley24}. Moreover, \citet{topping25} analyzed their results by comparing the log([\ion{N}{ii}]/H$\alpha$) ratio (as representative of metallicity), also finding a trend with EW(H$\beta$) and \Ne[\ion{S}{ii}]. Such a trend is also observed in Fig.~\ref{fig:stellar_age_properties} since any metallicity indicator that uses nitrogen lines depends on the N/O abundance. 

\begin{figure}
\begin{center}
    \includegraphics[width=0.45\textwidth, trim=0 0 0 0,  clip=yes]{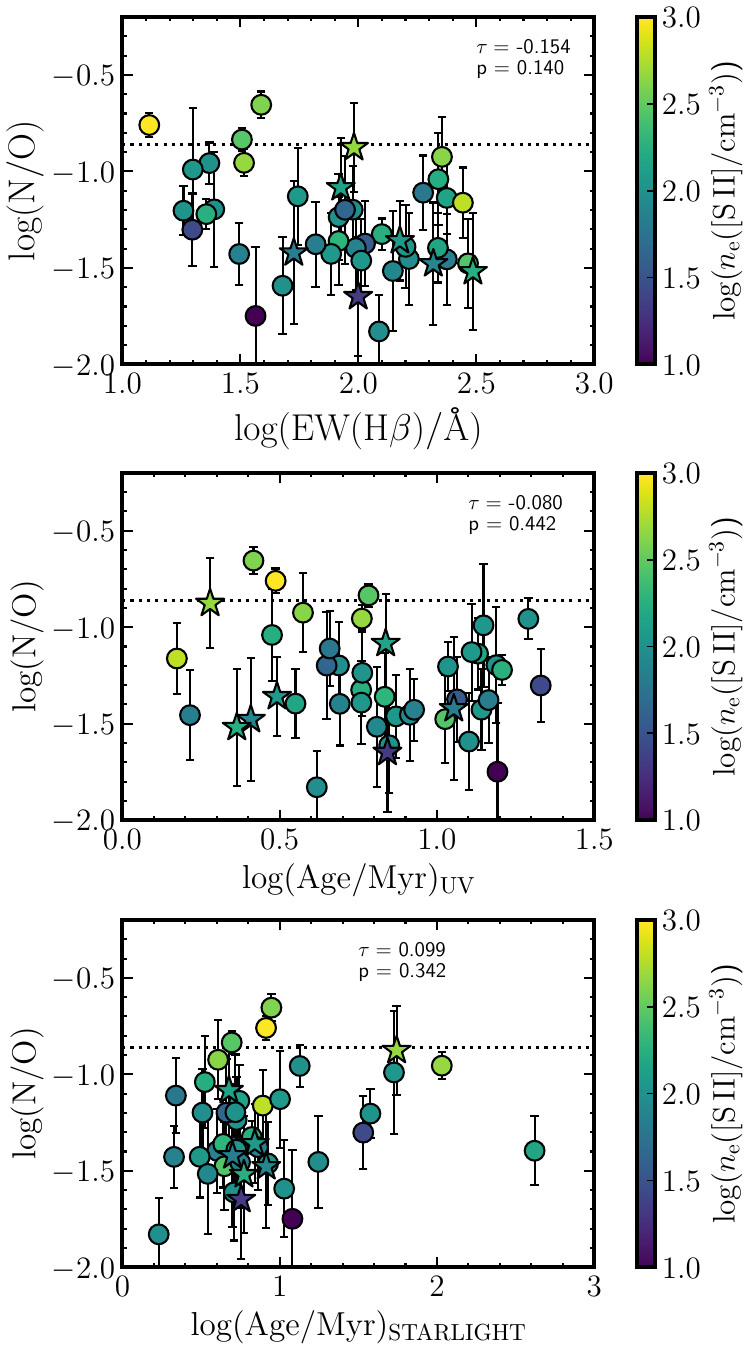}
        \caption{Comparison of EW(H$\beta$) and stellar age derived from the stellar population fitting from the UV and optical (see also Parker et al. 2025 and  James et al. 2025) as a function of the N/O ratio for CLASSY and in color-codded with \Ne. There is no statistical correlation between N/O and the EW(H$\beta$) or stellar age. The dotted lines represent the Solar value of N/O. The stars symbols indicate galaxies with WR features.}
\label{fig:stellar_age_properties}
\end{center}
\end{figure}

Following our analysis of EW(H$\beta$), we have used the stellar ages derived from the results of the best fit to the stellar populations presented in Parker et al. (2025) from the UV, which follows the procedure of \citet{chisholm19} (see also Sec.~\ref{sec:Archival_data}). 
The middle panel of Fig.~\ref{fig:stellar_age_properties} presents the analysis of N/O as a function of stellar ages derived from the young stellar population in the CLASSY galaxies. Based on the UV fits, the CLASSY galaxies show UV derived stellar ages ranging from 1 to 30 Myr. In our analysis, we do not find any significant correlation between N/O and UV-based stellar age (Kendall's coefficient of $\tau = -0.080$, $p$-value = 0.442). However, we find that high-gas-density galaxies (\Ne\ $> 300$ cm$^{-3}$) tend to show higher N/O values and younger ages, as also indicated by the analysis using EW(H$\beta$). In contrast, galaxies with lower densities (\Ne\ $< 200$ cm$^{-3}$) exhibit a more constant N/O value for a fixed age. It is important to note, that UV observations trace only the youngest stellar populations, while older stellar populations likely contribute significantly to the overall stellar age of the galaxies. 

We have also compared the stellar ages derived from stellar population modeling using {\sc Starlight} \citep[][]{cidfernandes2005} applied to the optical spectra from CLASSY \citep[][]{arellanocordova22a}, as derived in James et al. (2025), which may better trace the contributions from older stellar populations to nitrogen abundances. These authors estimate stellar ages by calculating the statistical weights of the stellar light contribution from the best-fit stellar population models obtained with {\sc Starlight}. The bottom panel of Fig.~\ref{fig:stellar_age_properties} shows the comparison between N/O and the stellar ages derived from this optical fitting. For most of the galaxies (approximately 80\%), the resulting ages are less than 10 Myr, while a smaller fraction show ages between 10 and 100 Myr. These optical derived ages tend to be higher than those obtained from UV fits and from EW(H$\beta$) due to the increased contributions of more evolved populations to the optical and the UV. While the youngest stellar populations in the UV are associated with the highest N/O, the light-weighted optical ages are appreciably higher, suggesting an important underlying older population capable of contributing enrichment via AGB stars. Interestingly, one of the WR galaxies, Mrk~996 (also known as J0127-0619), shows an older age ($\sim$50 Myr) compared to other WR galaxies, which is unexpected. It is likely that the stellar age of this galaxy is overestimated due to the complexity of its optical spectrum \citep[see][]{james09}. 

Overall, Kendall's coefficient ($\tau = 0.099$, $p$-value = 0.324) indicates no significant correlation between N/O and the stellar age derived from the optical fitting. Thus, although different stellar age indicators have their own caveats, we cannot discern a clear impact of stellar age on the N/O–O/H relation. Nevertheless, our results suggest that stellar age might have only a minimal effect on shaping and scattering the N/O–O/H ratio.

\section{Some comments on individual CLASSY galaxies}
\label{B}
\subsection{Gas kinematics and \Ne\ from different gas components}

 Some CLASSY galaxies exhibit different velocity components (narrow and broad) in certain Balmer lines and [\ion{O}{iii}]~$\lambda$5007, which can be used to analyze the physical properties of the gas in each component \citep[e.g.,][]{Hogarth20, mingozzi22, peng25}. For example, \citet{Hogarth20} performed a chemical analysis using high spectral resolution of the compact CLASSY galaxy, J1429+0643, finding high N/O ratios associated with two narrow components (log(N/O) $= -0.80 \pm 0.16$ and $-1.06 \pm 0.11$), while the gas in the broad component appears to be more enriched in oxygen (12 + log (O / H) = 8.58), but shows a typical log(N/O) of $-1.33 \pm 0.12$ (see Fig.\ref{fig:NO_highz}). For J1429+0643, we derive log(N/O) $= -1.11 \pm 0.19$ using the integrated SDSS profile, consistent within the uncertainties with the N/O values from the narrow components. In terms of $n_{\rm e}$, \citet{Hogarth20} derived \Ne[\ion{S}{ii}] = 200-490 cm$^{-3}$. However, such broad gas components (i.e., velocity dispersion, $\sigma=$~240-1800 km/s) might be associated with  outflows driven by supernovae (SNe) \citep{Hogarth20, Izotov07} or by radiation pressure, which becomes more efficient at higher metallicities. 

Recently, \citet{peng25} studied the physical conditions and metallicities of different kinematic components in 14 CLASSY galaxies. In general, they reported high \Ne values across various velocity components (see also Sec.~\ref{sec:density}), with no significant variations in metallicity, consistent with the findings of \citet{Hogarth20}. In our analysis, we are unable to derive abundances for the broad component due to the lack of detection of this component in $T_{\rm e}$-sensitive lines or in any lines suitable for determining $n_{\rm e}$ from [\ion{S}{ii}]. Nevertheless, this study highlights the importance of understanding the key physical conditions, galaxy properties, and kinematics that influence the determination of gas-phase abundances of nitrogen, and likely other elements. Follow-up analyses using higher resolution spectra of the outflow properties using high resolution spectra will be essential to disentangle additional chemical enrichment pathways for N/O at $z \sim 0$ (see  also Sec.~\ref{sec:discussion}).


\bsp	
\label{lastpage}
\end{document}